\documentclass[oneside, 10pt]{acmart}
\pdfoutput=1
\usepackage{soul}
\usepackage{color}
\usepackage{url} 
\usepackage{booktabs}
\usepackage{subfigure}
\usepackage{tabularx}
\usepackage{mathrsfs} 
\usepackage{array}
\usepackage{amsfonts}
\usepackage{amsmath}
\usepackage{amsthm}
\usepackage{enumitem}
\usepackage{graphicx}
\usepackage{eufrak}
\usepackage{hyperref}
\usepackage{color}
\usepackage{array}
\usepackage{caption}
\usepackage{multirow}
\usepackage{ragged2e}

\usepackage{pythonhighlight}
\usepackage{longtable}
\usepackage{hyperref}[colorlinks,linkcolor=blue]

\AtBeginDocument{%
  \providecommand\BibTeX{{%
    \normalfont B\kern-0.5em{\scshape i\kern-0.25em b}\kern-0.8em\TeX}}}

\acmBooktitle{ }

\begin{document}

\title{A Survey of Deep Learning Models for Structural Code Understanding}

\author{Ruoting Wu}

\email{wurt8@mail2.sysu.edu.cn}
\affiliation{%
  \institution{Sun Yat-sen University of China}
}
\author{Yuxin Zhang}
\affiliation{%
  \institution{Sun Yat-sen University of China}
}
\email{zhangyx355@mail2.sysu.edu.cn}
\author{Qibiao Peng}
\affiliation{%
  \institution{Sun Yat-sen University of China}
}\textbf{}
\email{pengqb3@mail2.sysu.edu.cn}
\author{Liang Chen}
\email{chenliang6@mail.sysu.edu.cn}
\authornote{Contact Author}
\affiliation{%
  \institution{Sun Yat-sen University of China}
}
\author{Zibin Zheng}
\email{zhzibin@mail.sysu.edu.cn}
\affiliation{%
  \institution{Sun Yat-sen University of China}
}


\renewcommand{\shortauthors}{Chen et al.}


\begin{abstract}
In recent years, the rise of deep learning and automation requirements in the software industry has elevated Intelligent Software Engineering to new heights. The number of approaches and applications in code understanding is growing, with deep learning techniques being used in many of them to better capture the information in code data. In this survey, we present a comprehensive overview of the structures formed from code data. We categorize the models for understanding code in recent years into two groups: sequence-based and graph-based models, further make a summary and comparison of them\footnote{We provide a paper collection about deep learning models and datasets for structural code understanding in https://github.com/codingClaire/Structural-Code-Understanding}. We also introduce metrics, datasets and the downstream tasks. Finally, we make some suggestions for future research in structural code understanding field.
\end{abstract}


\maketitle

\section{Introduction}

In the last several decades, deep learning has made remarkable achievements in various areas and permeated every aspect of human lives, especially in the domain of multi-media data processing such as image recognition, speech recognition, and natural language processing. With the booming development of deep learning techniques, as well as cooperatively increasing open-source code communities and automation requirements in the software industry, deep learning techniques began to be applied to more specific tasks in software engineering in recent years. 

Conventionally, source codes are considered as plain text sequences that may be understood using various existing approaches, such as deep learning approaches in neural language processing(NLP). However, when applied directly to source code, NLP approaches have the drawback of ignoring the code's structural information. When the code is learned merely as a sequence of plain text, syntactic and semantic information that is crucial to understanding the code, as well as the many relationships between program entities, may be overlooked.
Hence, a surge of works of understanding source code with structural information is proposed in recent years, which lead by the research of deep learning on sequences and graphs, such as Transformer\cite{bahdanau2014neural}, Graph Neural Networks\cite{Wu2019survey}. These techniques and their variants are developed to cope with various tasks in source code understanding including code representation and other downstream tasks. Although these methods have achieved some improvements, structural code understanding is still facing many challenges, which are identified and summarized as follows:

\begin{itemize}
	\item \textbf{Code Structural Modeling}: Since conventional language models feed the sequence of source code tokens as inputs, the structural information in the code is usually neglected. Therefore as result, a number of challenges arise about how to use structural information in code successfully, such as how to model structural information in code effectively and how to select effective structural information for specific downstream tasks.
	\item \textbf{Code Generic Representation Learning}: Much of the current research focuses on learning code representations for specific programming languages, making learning generic code representations a challenge. It's about how to learn language-independent code representations that get beyond programming languages' constraints. 
	\item \textbf{Code Task-specific Adaptation}: The following adaptations remain a challenge: how to choose and design specific architectures for downstream applications such as code generation and program repair, how to process datasets for task specifications, and how to adapt models in few-shots learning, transfer learning, and cross-linguistic scenarios.
\end{itemize}

In this survey, we present a comprehensive overview of structural learning techniques for code representation learning. In summary, our contributions can be listed as below:
\begin{itemize}
\item We introduce the structures in code data as well as the generating procedure, then give a summary of the downstream tasks for structural code understanding.
\item We propose a new taxonomy of deep learning models for structural code understanding based on the structures, which are categorized into sequence-based models and graph-based models.
\item We outline the challenges, the open problems, and future directions for structural code understanding.
\end{itemize}

The survey is organized as follows. In Section \ref{preliminary}, we give some basic introduction of the structures in code and how they are extracted from code data. In Section \ref{seq-models} and Section \ref{graph-models}, we separately introduce the models based on which structure they generally used, indeed, the sequence-based models and the graph-based models. We first give an overview of how they change the structures and then categorized them by the core models. In Section \ref{discussion}, we offer a discussion and comparison between the sequence-based models and graph-based models. The downstream tasks after conducting code representation is introduced in Section \ref{tasks}. We then summarize the related metrics and datasets in Section \ref{metrics_and_datasets}. We try to discuss some open research questions in Section \ref{open-problems}. Finally, we draw our conclusion in Section \ref{conclusion}.

\section{Preliminary}
\label{preliminary}


\subsection{Structures in Code}

\subsubsection{Overview}
\ 

First, we'll go through the basic structures that programs can produce. Table \ref{notation_table} summarizes the most commonly used notations. The following is a piece of code snippet in Python we used to illustrate the structures in code data.

\begin{python}
def add(a,b):
    x=0
    if(a>b):
        x=a-b;
    else:
        x=a+b;
    return x

res=add(1,2)
print(res)
\end{python}

\begin{table*}[t]
	\centering
	\caption{Summary of notations.}
	\label{notation_table}
	\scalebox{0.69}{
		\begin{tabular}{c|c||c|c||c|c}
			\hline
			\hline
			Symbol                     &
			Description                &
			Symbol                     &
			Description                &
			Symbol                     &
			Description                  \\

			\hline
			\begin{tabular}[c]{@{}c@{}} $P$ \end{tabular}  &
			\begin{tabular}[c]{@{}c@{}} Program \end{tabular}  &
			\begin{tabular}[c]{@{}c@{}} \\ $S$\\ \\ \end{tabular}  &
			\begin{tabular}[c]{@{}c@{}}  Code Snippet \end{tabular}  &
			\begin{tabular}[c]{@{}c@{}} \\ $F$\\ \\ \end{tabular}  &
			\begin{tabular}[c]{@{}c@{}} function  \end{tabular}
			\\
			\hline
            \begin{tabular}[c]{@{}c@{}} \\ $A$\\ \\ \end{tabular} &
            \begin{tabular}[c]{@{}c@{}}  Abstract Syntax Tree of a code snippet \\ or a program \end{tabular} &
            \begin{tabular}[c]{@{}c@{}} $G_c$ \end{tabular} &
            \begin{tabular}[c]{@{}c@{}} Control flow graph of a code snippet \\ or a program\end{tabular} &
            \begin{tabular}[c]{@{}c@{}} $G_d$ \end{tabular} &
            \begin{tabular}[c]{@{}c@{}} Data flow graph of a code snippet \\ or a program\end{tabular} 
            \\

            \hline
            \begin{tabular}[c]{@{}c@{}} $path$ \end{tabular} &
            \begin{tabular}[c]{@{}c@{}} Sequence of nodes extracted \\
            from the AST\end{tabular} &
            \begin{tabular}[c]{@{}c@{}} \\ $n$\\ \\ \end{tabular} &
            \begin{tabular}[c]{@{}c@{}}
            The length (token number) of a code snippet \\ or a program
            \end{tabular} &
            \begin{tabular}[c]{@{}c@{}} \\ $y$\\ \\ \end{tabular} &
            \begin{tabular}[c]{@{}c@{}} Labels of the code \end{tabular}
            \\

            \hline
            \begin{tabular}[c]{@{}c@{}} $D$ \end{tabular} &
            \begin{tabular}[c]{@{}c@{}} Code description in natural language \\
            from the AST\end{tabular} &
            \begin{tabular}[c]{@{}c@{}} \\ $t$\\ \\ \end{tabular} &
            \begin{tabular}[c]{@{}c@{}} single token \end{tabular} &
            \begin{tabular}[c]{@{}c@{}} \\ $H$\\ \\ \end{tabular} &
            \begin{tabular}[c]{@{}c@{}} the intermediate representation \\ of the code \end{tabular}

\\

\hline
			
			\hline
			\hline
		\end{tabular}
	}
\end{table*}

\begin{figure}[h] 
\centering
\includegraphics[width=1\textwidth]{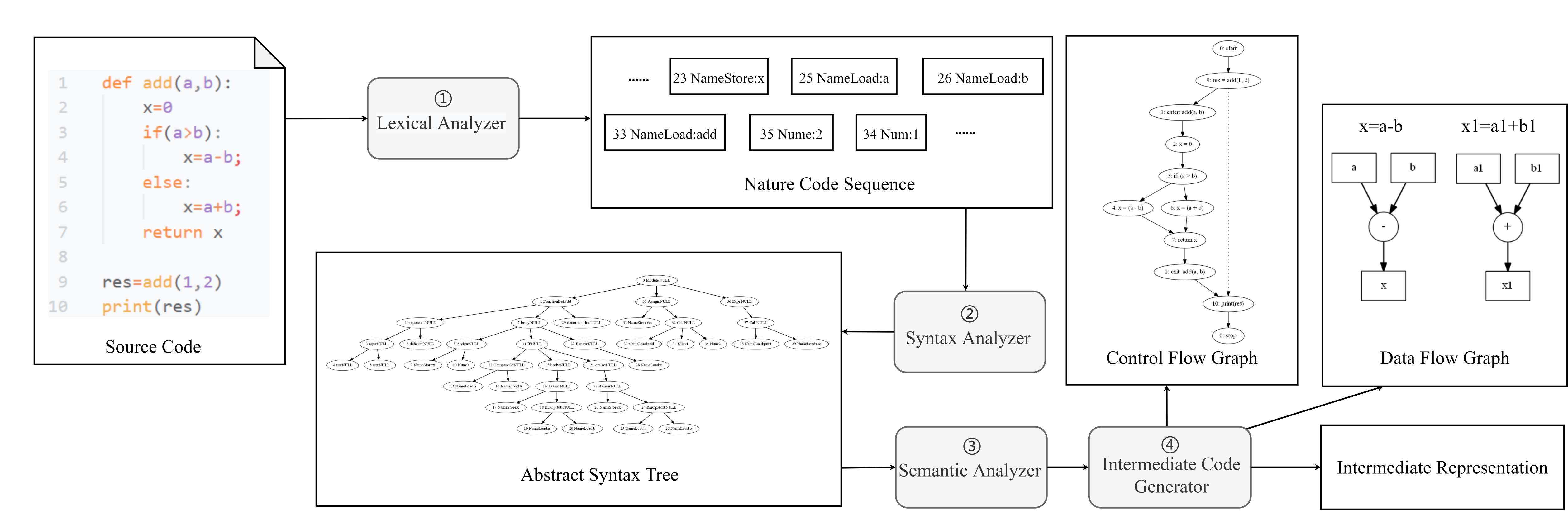} 
\caption{The basic structures generated from code includes (a) Nature Code Sequence, (b)  Abstract Syntax Tree, (c) Control Flow Graph and (d) Data Flow Graph. The front end of a compiler constitutes the four components (Lexical Analyzer, Semantic Analyzer, Syntax Analyzer, and Intermediate Code Generator)  } 

\label{pic:structure} 
\end{figure}

The general techniques for generating structures from the source code snippet are shown in Fig. \ref{pic:structure}. The Lexical Analyzer first converts the code into a token-based sequence. Each token in the sequence has two attributes, type and value. Lexical Analysis of code is similar to the tokenization stage in natural language. Inspired by Hindle et al.~\cite{Hindle2012OnTN}, we refer to these unprocessed code sequences as \textit{Natural Code Sequence (NCS)} in this article for the consistence of discussion, which may be called as token sequence or code sequence in other articles. 

The Syntax Analyzer, also known as a parser, takes the tokens and produces an \textit{ Abstract Syntax Tree(AST)} based on the code snippet's grammar. The Abstract Syntax Tree is then utilized in the Semantic Analyzer to verify for semantic consistency, and the Intermediate Code Generator to construct the Intermediate Representation, which varies depending on the programming language. Control flow and Data flow are both graph-like Intermediate Representations known as \textit{Control Flow Graph(CFG)} and \textit{Data Flow Graph(DFG)}, respectively.

Other structures established in software engineering, such as UML Graph and Program Dependency Graph, are in addition to the basic structures.

\subsubsection{Nature Code Sequence}
\label{NCS}
\ 

Given a Program $P$ or a  Code snippet $S$,  the  Nature Code Sequence(NCS) $S=\{t_1,t_2,... ,t_n\}$ is obtained by Lexical Analyzer, $t_i$ refers to the token in the code. Using NCS to represent the code is the simplest and most common approach. The position of the token in the sequence corresponds to the order in which it appears in the code snippet.

\subsubsection{Abstract Syntax Tree}
\label{AST}
\

\begin{figure}[h] 
\centering
\includegraphics[width=1\textwidth]{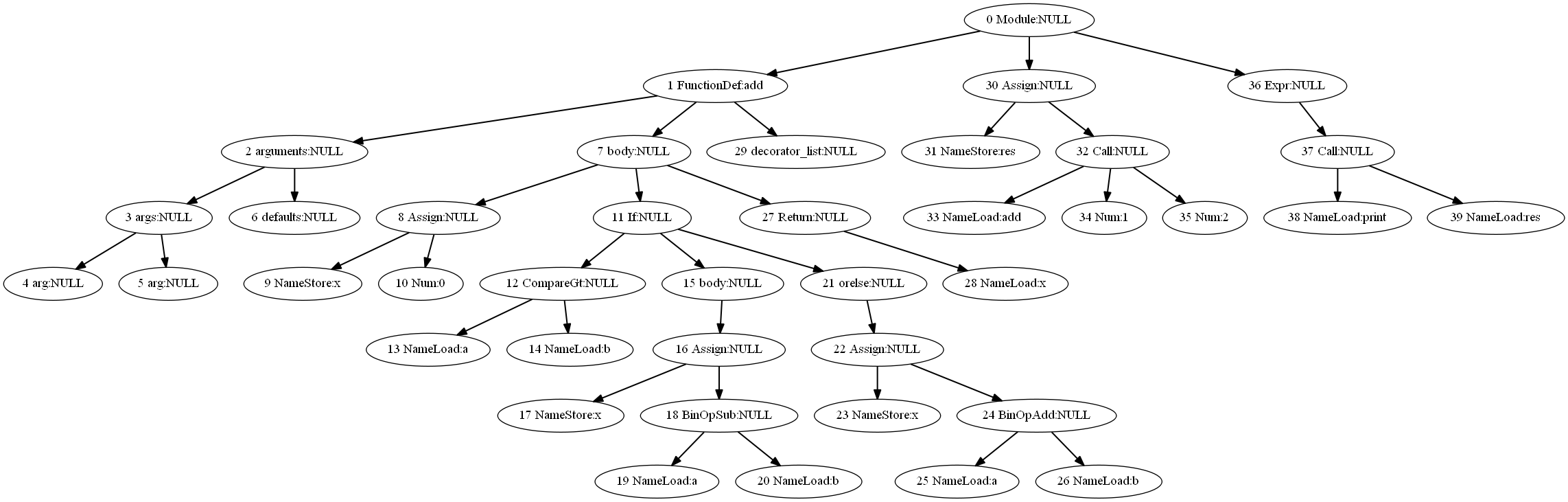} 
\caption{An example of  Abstract Syntax Tree} 
\label{pic:ast} 
\end{figure}

Given the language grammar, the Abstract Syntax Tree (AST) of code is generated by Syntax Analyzer and marked as $A$, which hierarchically reflects the structural and syntax information of code. The root node of the syntax tree represents the start symbol. The interior nodes are the nonterminals in the grammar, while the leaf nodes are the terminals, which are usually the variables and identifiers defined by the programmers from the NCS. Fig. \ref{pic:ast} is the AST of the above code snippet in Python.

\subsubsection{Flow Graph}
\label{flow_graph}
\begin{figure}[h] 
\centering
\includegraphics[width=0.3\textwidth]{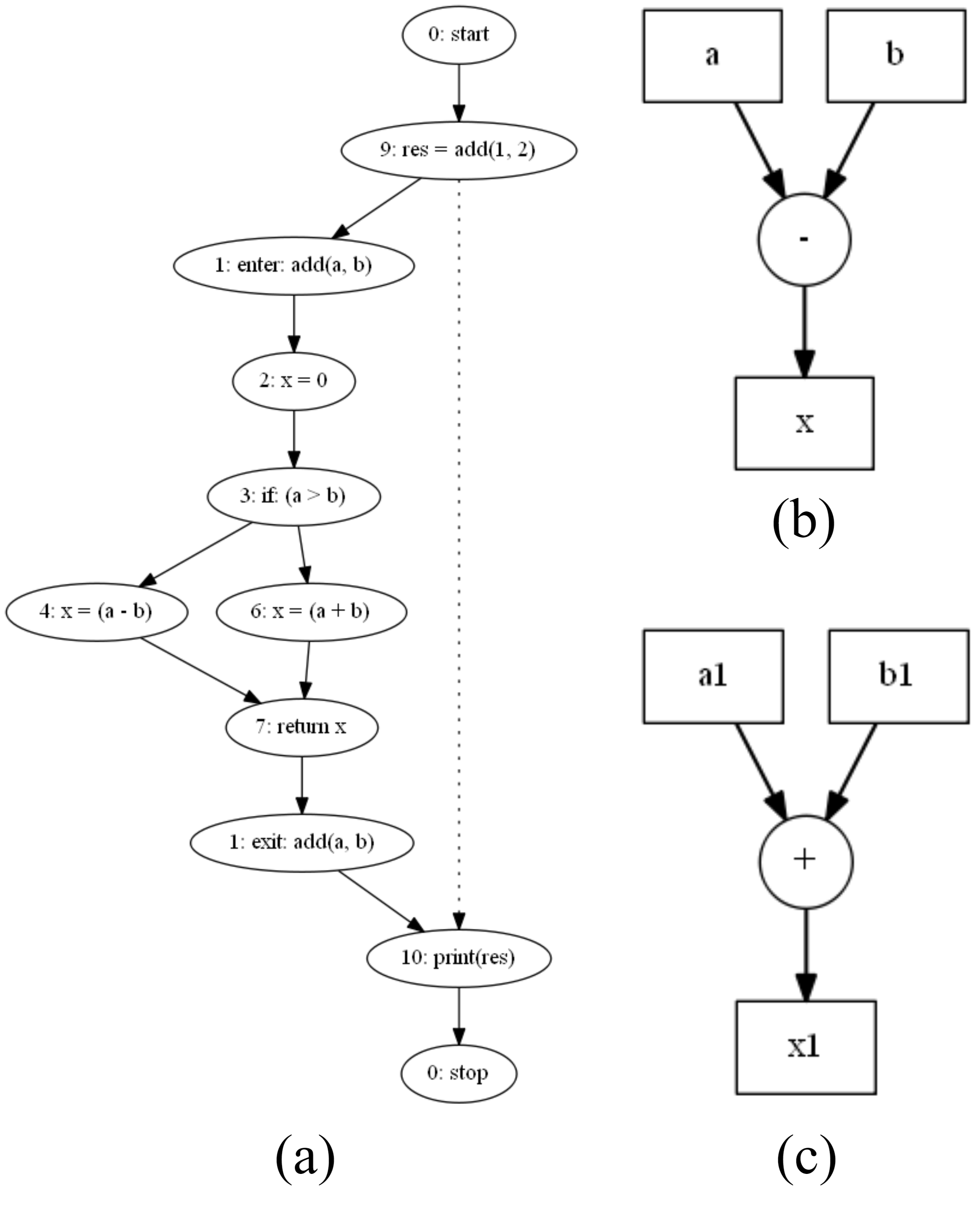} 
\caption{An example of  Flow Graphs. (a) Control Flow Graph, (b) Data Flow Graph of expression $x=a-b$, (c)Data Flow Graph of expression $x1=a1+b1$.} 
\label{pic:flow_graph} 
\end{figure}

\
Flow Graphs are the graphs that cover the semantic information of source code. The two typical flows in flow graphs are control flow and data flow, which separately represent how the program executes and how the data flows. Fig.\ref{pic:flow_graph} shows the example of control flow and data flow graphs for the python code snippet mentioned.


\paragraph{\textbf{Control Flow Graph}}
The Control Flow Graph (CFG) of a program marked as $G_c$, represents different execution paths of a program. Each node of CFG is a basic block that represents a behavior without branches. The edges of the graph represent how these basic blocks flow from one to another. In the example code snippet, the judgement of $a>b$ will cause the program into 2 branches, one is $x=a-b$, the other is $x=a+b$. As the Fig.\ref{pic:flow_graph} (a) , the CFG has two edges from the decision node(if-statement) to the two downward nodes(two basic blocks). 

\paragraph{\textbf{Data Flow Graph}}
The Data Flow Graph (DFG) of a program marked as $G_d$ represents the dependency relation between variables. DFG can represent code snippets without conditionals. In the example code snippet, we select two statements: $x=a-b$ and $x=a+b$ to draw their DFGs. To eliminate the repeated assignment to $x$, we rename variables in the second assignment, convert them to $x1=a1+b1$. Therefore, the two DFGs are shown in  Fig.\ref{pic:flow_graph} (b) and (c). Each data flow node represents the operation of variables and each edge represents how the value flows.

\paragraph{\textbf{Control/Data Flow Graph}}
Because the DFG can only represent basic blocks without branches, it can be used to replace the basic blocks of a CFG, resulting in a Control/Data Flow Graph (CDFG). There are two types of nodes in a program's CDFG, the decision node and the data-flow node.


\subsubsection{Other Structures}
\label{other structure}
\

In addition to the above structures, there are some other code structures that are less common in code understanding.

\paragraph{\textbf{Other Intermediate Representation}}
Intermediate Representation(IR) is a data structure that can be obtained from a compiler such as LLVM Compiler Infrastructure\cite{lattner2004llvm}. The frontend Compiler compiles the source code and generates an IR for the backend Compiler to optimize and translate. In LLVM infrastructure, the IR is in Static Single Assignment(SSA) form. The Broad definition of IR includes the flow graphs as well as other graph structures. Program Dependency Graph(PDG)\cite{ferrante1987program} is one of the intermediate representations that make explicit both data and control dependencies for each operation in a program. PDG are useful to perform optimizations through a single walk.

\paragraph{\textbf{The Unified Modeling Language}}
The Unified Modeling Language (UML) is a widely-used language for specifying, visualizing, and documenting the artifacts of a software-intensive system. UML class diagram is a type of static structure diagram that describes the structure of a system by showing the system classes and their attributes, operations (or methods), and relationships.
 

\subsection{Deep learning models}
In this section, we focus on deep learning models commonly used in code understanding tasks that have shown to be effective in other domains and are now being used by a growing number of researchers in the code domain.
\subsubsection{Recurrent Neural Network}
\label{RNN}
\

Recurrent Neural Networks (RNNs)\cite{rumelhart1986learning} are the neural networks where each unit is connected by directed cycles. RNNs can use their hidden state to track the long-term information of the sequence data. Therefore, RNNs are a common choice for sequence modeling. Vanilla RNNs have gradient vanishing and gradient explosion  problems, which can reduce the ability of a model to learn long-term information. Long Short-Term Memory(LSTM) and Gated Recurrent Units(GRU) are the two most used RNN models that can avoid the problem and achieve better results.

 Long Short-Term Memory(LSTM)\cite{LSTM} has basic model from RNNs. Every unit of LSTM considers the hidden state, the current input, and the information from its memory cell. LSTM uses 3 gates, input gate, forget gate and output gate to control the learning and transfer of information. 

Gated Recurrent Units(GRU) \cite{cho2014learning} combine the input gate and forget gate in LSTM into one gate, called the update gate, and the other gate is the reset gate. The reset gate and the update gate can control the degree of memory or forgetting of sequence information by the hidden state. Compared with LSTM, GRU has comparable performance but lower computational cost.

\subsubsection{Transformer}
\label{Transformer}
\ 

Bahdanau et al.\cite{bahdanau2014neural} propose the Attention mechanism to solve the problems of excessive information length and information loss in machine translation tasks. It feeds all of the hidden states from the Encoder into the Decoder after linear weighting and assigns various attention weights to each input token, indicating which inputs are more significant to the output. 

To enhance computational performance and better describe global relationships in sequences, Vaswani et al.\cite{vaswani2017attention} propose self-attention. It is a special attention mechanism, so that information from any position in the sequence can directly affect the encoding of the other token. Based on self-attention they propose a new neural network model called Transformer which consists of multiple attention blocks made up by self-attention. Transfomer's encoder uses the self-attention mechanism to associate the tokens in the input sequence with all other tokens as it learns the representation of the input. Also, the input of the Transfomer's decoder is associated with the output of the encoder through self-attention. Transformer and its variants, such as Bert, GPT, etc., are capable of processing complex data and can process large amounts of sequence data. Therefore they are often used as pretraining models to capture the rich information from large amounts of complex data. 

\subsubsection{Graph Neural Network}
\

Graph Neural Networks(GNNs) are deep learning models using message passing between nodes to capture structural information and aggregate semantic information in graphs. GNN can be categorized into four groups according to the survey by Wu et al.\cite{Wu2019survey}: Recursive GNNs, Convolutional GNNs, Graph autoencoders, and spatial-temporal GNNs. GNNs are widely used in node classification, edge prediction, and graph classification tasks. 

The following are the typical models used in code representation-related tasks. Gated graph Neural Network(GGNN) \cite{Li2016} uses gated recurrent units and unroll the propagation process for a fixed number of timesteps. The representations of nodes are the final step output. Graph Convolution Network(GCN) \cite{kipf2017semisupervised} is one of the convolution GNNs, which stack multiple graph convolutional layers to better extract the information from neighbors. Graph Attention Network(GAT)\cite{Velickovic2018} uses the attention mechanism in the message-passing step to aggregate neighborhoods' information with different weights and update the encoding of the node. The above Graph Neural Networks are typical deep learning models for learning the representation of graphs or nodes in code data.

\subsubsection{Encoder-Decoder Framework}
\

The Encoder-Decoder framework\cite{cho2014learning} is presented as a solution to traditional machine translation difficulties. The input language is encoded in the encoder section to obtain the intermediate representation Context. And then in the decoder part, corresponding outputs are generated one by one based on Context and related inputs. Sutskever et al.\cite{sutskever2014sequence} present the seq2seq model based on Encoder-Decoder framework to overcome the problem of indefinitely long input-output sequences, which aids sequence output with special markers such as <Eos>. Different encoders and decoders can be selected according to particular tasks, such as RNN-based models, Transformer-based models, GNN-based models, etc. Encoder-decoder model architecture has become the mainstream approach to address the code generation issue and other tasks due to the naturalness and sequence of code. For example, Rabinovich et al.\cite{MaximRabinovich2017AbstractSN} introduce a syntax network (ASN) that extends the encoder-decoder framework to generate AST. 




\section{Sequence-based Models}
\label{seq-models}
Sequential models that perform well in sequence-related tasks, such as the Recurrent Neural Network family\cite{rumelhart1986learning, LSTM,cho2014learning} and Transformer\cite{vaswani2017attention}, can be effectively applied to code-related tasks. They can be used to encoder-decoder architecture to fulfill downstream tasks such as code summarization and code generation, as well as learn to access code representation for downstream activities.

The frequently used term \textit{code sequence} refers to natural code sequences created by the code itself, which is referring to NCS introduced by section \ref{NCS}. Because the code is highly structured, there are sequences formed by pre-processing the code structure input such as AST introduced in section \ref{AST}, which refer to the \textit{flattened sequence}. In this section, we introduce the models for processing serialized code data mentioned above. We first show the structural transformation of code including the different methods of pre-processing the structures of code to get the flattened sequence, and then show the models for processing NCS and flattened sequences respectively.
\subsection{Structure Transformation}
\label{seq-trans}
\begin{figure}
	\centering
	\includegraphics[width=160mm]{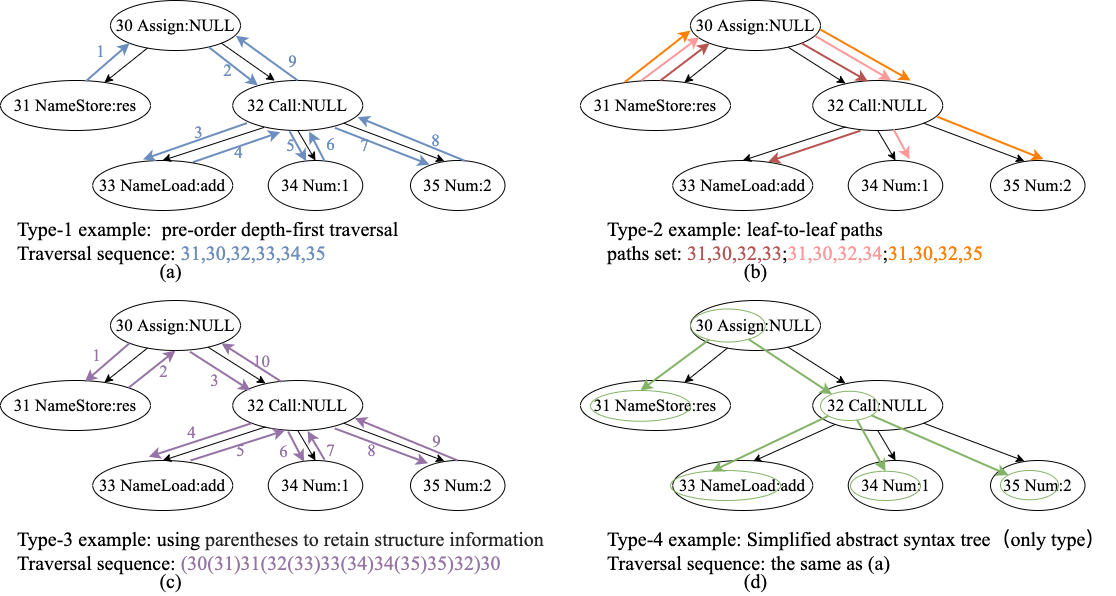}
	\caption{Examples of local AST using four types of structure transformation to obtain flattened sequence, (a) using depth first traversal to convert AST into sequence, (b) showing three paths in the leaf to leaf path set, (c) showing SBT\protect\cite{Hu2018DeepCC} using parentheses to retain structure information, (d) only retaining the type information of nodes in AST, and the obtained sequence is the same as (a).}
	\label{Fig1}
\end{figure}

The AST format is commonly used to express source code structural information. For the sequence model to make efficient use of the code's structural information, some strategies for flattening the AST are offered. Flattening procedures are divided into four types, as shown in Fig \ref{Fig1}, which shows the varied sequences obtained by different types of flattening approaches.
\paragraph{\textbf{Type-1: Depth-first traversal}}  Because the AST represents the code's structural information as a tree, traversing the tree structure depth-first, so that the nodes on each subtree are adjacent in the sequence, is the simplest way to extract the flattened sequence. More models advocate preorder depth-first traversal because the root node (operator) of each subtree in the AST is frequently the subtree's center. Type-1 refers to the Structure Transformation method that uses a depth-first traversal on the AST.

\paragraph {\textbf{Type-2: AST path}} The approach of serializing the AST via paths is recommended because every two nodes in the AST have a path between them. Methods for extracting paths from ASTs include paths between arbitrary nodes, paths between terminal nodes, paths from terminal nodes to root nodes, and so on. In code generation task, paths from terminal nodes to new nodes are also used. Type-2 refers to the approach of Structure Transformation that uses paths between nodes on the AST.
\paragraph{\textbf{Type-3: Structure information addition}} To better keep structural information in the AST and make the flattened sequence unique, Type-1/2-based structure information adding methods are proposed. The Structure-based traversal approach, for example, uses brackets to represent the AST structure, and the brackets in the created sequence may be used to detect the subtree of a certain node, allowing the generated sequence to be translated back to the AST. The use of "<" and ">" to enclose the non-terminal node's subtree, which corresponds to the code block, is another technique to preserve structural information. As a result, we refer to the Structure Transformation method as Type-3 because it incorporates code structure information in the obtained sequence.

\paragraph {\textbf{Type-4: AST partial retention} } AST contains a lot of structural and syntactic information about the code, but it also contains a lot of useless data. As a result, numerous ways suggest filtering the ASTs, preserving the nodes of interest, and then flattening the sequence derived from the filtered ASTs. To generate flattened sequences in code defect detection jobs, only three types of AST nodes are generally kept, for example, method call and class instance creation nodes, declaration nodes, and control flow nodes. Type-4 refers to the Structure Transformation approach, which keeps only important nodes. 

\subsection{Natural code sequence}
\ 
NCS is a natural way to represent whole code fragments because code is made up of individual tokens. The approaches and models for comprehending code using NCS are described in this section.

\paragraph{\textbf{N-gram Models}} N-gram models, widely used as early language processing approaches, assume a Markov property that the probability of the current word is only affected by its prefixed N-1 words, which can capture the statistical characteristics of a sequence to some extent. In order to take advantage of the statistical properties of NCS, some early models using N-Gram models to accomplish code representation tasks were proposed.
Hindle et al.~\cite{Hindle2012OnTN} firstly adopt N-gram models on NCS and find that the language models are conducive to extracting local statistics by exploring the naturalness rather than the syntax or semantics. 
Tu et al.~\cite{Tu2014OnTL} cooperate N-gram model with a cache component to further capture the localness of source codes.
Karaivanov et al.~\cite{Karaivanov2014PhraseBasedST} exploit the N-gram model for phrase-based programming language translation.

The N-gram model takes all of the information from the N-1 tokens before the token to learn its representation, but it cannot use remote token information (such as the reuse of a variable in the code), and it also cannot build identical vector representations for tokens with similar meanings. As a result, the N-Gram model was gradually replaced by the more learning model introduced later in the code representation task.

\paragraph{\textbf{Convolutional Neural Network}} Convolutional neural network (CNN) is first proposed for processing images to learn image representations by capturing features of local images through convolutional kernels. Also, It can extract significant n-gram features from input sentences to create a semantic representation of the potential of sentence information for downstream tasks, while effectively capturing rich structural patterns in the code, so some work on learning code representations with CNN was proposed. 
To summarize the code snippet, Allamanis et al.~\shortcite{Allamanis2016} employ an attentional neural network that uses convolution on the input tokens to detect local time-invariant and long-range topical attention features in a context-dependent way.
For better code searching, CARLCS-CNN\cite{Shuai2020ImprovingCS} first embeds code and query respectively using CNN since CNN can capture the informative keywords in query and code, then learns interdependent representations for the embedded code and query by a co-attention mechanism. CNN is unable to model long-range dependencies in code sequences, there are local limitations on the sensitivity to word order, and since code sequences are often large, CNN models are less often used in practical applications to learn to understand code.

\paragraph{\textbf{Recurrent Neural Network}} 
As previously section \ref{RNN} stated, RNN and their variants perform exceptionally well on sequential tasks, resulting in a huge number of RNN-based models for code-related tasks to handle NCS.
Veselin et al.~\cite{Veselin2014CodeCW} have shown that the well-trained RNN can outperform N-gram models~\cite{Hindle2012OnTN} when processing NCS.
Dam et al.~\cite{dam2016deep} propose to use LSTM to predict the next token in order to address the inability of n-gram models to capture token dependencies in sequences. 
CodeNN~\cite{iyer2016summarizing} uses LSTM with attention to produce sentences that describe code snippets.
Liu et al.\cite{Liu2016LatentAF} employ latent attention over outputs of a Bi-LSTM to better translate natural language descriptions into If-Then program. 
Bhoopchand et al.\cite{AvishkarBhoopchand2016LearningPC} enhance LSTM with a pointer network specialized in referring to predefined classes of identifiers to well capture long-range dependencies in the code, thus giving better suggestions for the next token input.
CODEnn~\cite{Gu2018DeepCS} provides a deep architecture composed of a code embedding network, description embedding network, and a similarity module to align the embeddings of code-description pairs. The code embedding network and description embedding network are both use LSTM. 
Tal et al.~\cite{Tal2018NeuralCC} utilize RNNs to learn a language-agnostic intermediate representation that is generated from code syntactical structures.
Vasic et al.\cite{MarkoVasic2019NeuralPR} present a solution to the general variable-misuse problem in which enumerative search is replaced by a neural network containing LSTM that jointly localizes and repairs faults.
CodeGRU~\cite{Hussain2020CodeGRUCD} further applies GRU in code sequence processing to capture contextual dependencies.
Compared with CNN, RNN family can handle arbitrary length input and has more flexible code sequence modeling ability, but vanilla RNN has the problem of gradient disappearance and gradient explosion. LSTM and GRU, as variants of vanilla RNN, can learn long-term dependency and solve the problem of gradient explosion and disappearance to a certain extent, and gradually become the current RNN family Core.
\paragraph{\textbf{Transformer}}
Transformer can successfully handle the distant dependence problem, overcome the limitation that RNNs cannot be computed in parallel, and employ self-attention to generate more explanatory models, as described in section \ref{Transformer}. Transformer is being used for a growing number of sequence-related work, and NCS is no exception.
Ahmad et al.~\cite{ahmad2020transformer} first employ the Transformer for code summarization to handle the ubiquitous long-range dependencies in source code from natural code sequence.
TFix\cite{BerkayBerabi2021TFixLT} works directly on program text and phrases the problem of code fixing as a text-to-text task, so it can leverage a powerful Transformer based model pre-trained on natural language and fine-tuned to generate code fixes. 
CodeBERT~\cite{Feng2020CodeBERTAP},CuBert\cite{Kanade2020}, GPT-C\cite{svyatkovskiy2020intellicode} and CodeT5\cite{YueWang2021CodeT5IU} use both NCS and related natural language to pre-train the Transformer architectures for downstream tasks, such as code search, code clone detection and code summarization. OSCAR\cite{Peng2021}  and GraphCodeBERT\cite{guo2020graphcodebert} are also pre-trained models based on transformer using NCS while exploiting the semantic information of flow graph.  OSCAR adds GCF information to the model training using positional encoding. GraphCodeBERT takes DFG as part of the input while exploiting the node and edge relationships for Graph-Guided Masked Attention to better understand the code.

Transformer can effectively learn a big quantity of data, but its memory and processing requirements are enormous when compared to models like RNN. Following that, work should be measured in terms of resource consumption and performance improvement, and a suitable model should be chosen by taking into account the current circumstances.

\paragraph{\textbf{Others}}
Other models have been incorporated in related work in addition to CNN, RNN, and other models that are extensively utilized in NCS related tasks.
Sachdev et al.\cite{Sachdev2018RetrievalOS} create a continuous vector embedding of each code fragment at method–level granularity, map the given natural language query to the same vector space, and use vector distance to simulate relevance of code fragments to a given query.
CCLearner\cite{li2017cclearner} extracts token sequence from known method-level code clones and non-clones to train a deep Neural Network(classifier) and then uses the classifier to detect clones in a given codebase.
SCC\cite{KamelAlreshedy2018SCCAC} is a classifier that can identify the programming language of code snippets written in 21 different programming languages, it employs a Multinomial Naive Bayes(MNB) classifier trained using Stack Overflow posts.
Sachdev et al.\cite{Sachdev2018RetrievalOS} propose a simple yet effective unsupervised model that combines word2vec\cite{mikolov2013distributed} and information retrieval methods for code search.
UNIF~\cite{Cambronero2019WhenDL} first uses word2vec to embeds code/quey and then combines code embedding with attention. These models show better results in specific tasks, and therefore, subsequent work dealing with NCS should not be limited to the adoption of mainstream models.

\subsection{Flattened Sequence}
The four structural transformations indicated in \ref{seq-trans} can be used to obtain flattened sequences. Although the flattening procedure consumes more resources, the flattened sequences preserve some structural information about the code, and the applicable models can learn more about the code from the flattened sequences than NCS. 
The sequences obtained by different structural transformations may be suitable for different models. The models for processing the flattened sequence are described in the following and these models also show good performance in the natural language processing field. 
We make a finer segmentation of the RNN family in this section, including Vanilla LSTM, Bi-direction LSTM, and GRU, because of the enormous variety of applications of RNNs and their variants on flattened sequences.
\paragraph{\textbf{Word2vec}}word2vec is a method of converting an input token into a vector representation, where the converted vector contains, to some extent, the contextual information of the token. Word2vec contains two training models, CBOW (Continuous Bag-of-Words Model) and Skip-gram (Continuous Skip-gram Model), which have strong generality and are therefore used in early work on flattened sequence for learning code representation.
API2Vec\cite{TrongDucNguyen2017API2vec} traverses the AST using \textbf{Type-4} to build an annotation sequence according to the syntactic units related to APIs. These sequences are then used to train CBOW to generate API embeddings that may be utilized to migrate equivalent API usage from Java to C\#.
Alon et al.\cite{UriAlon2018AGP} first use \textbf{Type-3} to obtain flattened sequence by adding up and down momentum information to the paths between nodes and then use word2vec to complete the prediction of method names. 
Because Word2vec is unable to learn the representation of polysemous words and cannot successfully capture long-range dependencies, further work is being done to combine Word2vec with other models to learn flattened sequences better.
\paragraph{\textbf{Deep Belief Network}} Deep Belief Network(DBN)\cite{YoshuaBengio2009LearningDA} is a generative model which uses a multi-level neural network to learn a representation from training data that could reconstruct the semantic and content of input data with Maximum probability. DBN can be used to identify features, classify data, generate data, and do other tasks. 
Wang et al.\cite{SongWang2016AutomaticallyLS,SongWang2020DeepSF} use \textbf{Type-4} and then use DBN to complete the defect prediction. They produce flattened sequences from only three sorts of AST nodes: method invocation and class instance creation nodes, declaration nodes, and control flow nodes. Because the names of methods, classes, and types are usually project-specific, there are very few methods with the same name across multiple projects. To get better detection results, they extract all three classes of AST nodes for cross-project defect prediction(CPDP), but instead of utilizing their names, it uses their AST node type, such as method declaration and method invocation, for declaration nodes and control flow nodes.
DBN can automatically learn semantic features from token vectors extracted from ASTs and is one of the first non-convolutional models to be successfully trained by applying deep architectures. However, DBN has mostly lost favor and is rarely used compared to other unsupervised or generative learning algorithms nowadays.

\paragraph{\textbf{Vanilla Long Short-Term Memory}}
Vanilla LSTM is the LSTM introduced in section \ref{RNN}, which is capable of learning long-range dependencies in some extent and is heavily used in work dealing with flattened sequences.
Liu et al.\cite{ChangLiu2017NeuralCC} use \textbf{Type-1} and explore several variants of simple LSTM architecture for different variants of the code completion problem.
Li et al.\cite{2017pointerMixtureNetork} utilize \textbf{Type-3}, which allows storing two extra bits of information about whether the AST has children and/or right siblings in the type node. And the pointer mixture network proposed consists of two main components: a global RNN component(LSTM) and a local pointer component, which utilizes the pointer network to point to the previous position in the local context according to the learned position weights to solve the OoV problem.
DeepCom \cite{Hu2018DeepCC} uses \textbf{Type-3} and design a new structure-based traversal(SBT) method to better preserve structural information in the code. The SBT traversal method uses parentheses to indicate the structure of the AST, and the parenthesis in the created sequence may be utilized to determine the subtree of a given node, allowing the generated sequence to be transformed back to AST. DeepCom employs the seq2seq model which uses LSTM as encoder and decoder to generate code fragment summaries. 
code2vec~\cite{Alon2019code2vec} employs \textbf{Type-2} to represent code fragments with the set of paths between all terminal node pairs in the code to complete the prediction of method names of the code and learns the representation of the sequence of internal non-terminal nodes using LSTM.
For better program classification, Compton et al.\cite{RhysCompton2020EmbeddingJC} investigate the effect of obfuscating variable names during the training of a code2vec model to force it to rely on the structure of the code rather than specific names and consider a simple approach to creating class-level embeddings by aggregating sets of method embeddings.
Seml\cite{liang2019seml} uses \textbf{Type-4} and sends the sequence obtained after filtering the AST to the LSTM to complete the defect detection of the code. 
\textbf{Type-3} is used by SA-LSTM\cite{FangLiu2020SALSTM}, which surrounds the sub-tree of each non-leaf node, which corresponds to code blocks, with < and >. SA-LSTM enhances the LSTM network with a stack to store and recover contextual information based on the code's structure for modeling the hierarchical structure of code.

Different works will modify the LSTM to suit their own tasks, but the LSTM has a disadvantage in parallel processing and cannot fully solve the gradient problem, as well as cannot do anything for a very large order of magnitude sequences, and Bi-LSTM and GRU introduced later also face the same problem. The latest effort will use LSTM or other RNN variants as a component of the model and mix it with other models to accomplish the task, in order to better exploit the advantages of RNN and its variants on sequence processing.
\paragraph{\textbf{Bi-direction Long Short-Term Memory}}Compared with the traditional LSTM which only retains the previous information, the bidirectional Long short Memory (Bi-LSTM) can also use the later information, which can better capture the semantic dependencies in both directions.
code2seq~\cite{Alon2018code2seq} employs the same \textbf{Type-2} as code2vec to obtain the flattened sequence, and Bi-LSTM was used to learn the representation of the internal non-terminal nodes sequence.
DeepCPDP\cite{DeyuChen2019DeepCPDPDL} uses \textbf{Type-4}, using simplified Abstract Syntax Tree(SimAST) to represent the source code of each extracted program. DeepCPDP uses SimASTToken2Vec, an unsupervised-based embedding approach, and will classify the code inputted as defective or non-defective using Bi-LSTM with attention mechanism and Logistic regression.
Pythia \cite{Alexey2019Pythia} uses \textbf{Type-1} to predict the method names and API calls that the developer wants to use in programming. Pythia aggregates the initial(obtained by word2vec) and intermediate(learned by Bi-LSTM) vector representations through a fully connected layer to obtain the final vector for prediction.

\paragraph{\textbf{Gated Recurrent Units}}
The GRU introduced in section \ref{RNN} is a simplified version of LSTM with a reduced number of gates and therefore easier to converge with relatively few parameters, so some work will choose to use GRU to learn the representation of the flattened sequence.
ast-attendgru\cite{AlexanderLeClair2019ANM} uses \textbf{Type-3} and proposes SBT-AO which modifies the SBT AST Flastting procedure to simulate the case when only an AST can be extracted. SBT-AO replaces all words (except official Java API class names) in the code to a special <OTHER> token, remaining all the code structure in SBT. It uses two GRU with attention mechanism to process NCS and SBT-AO respectively for getting context vector and then predicts the summary one word at a time from the context vector, following what is typical in seq2seq models.
Hybrid-Deepcom\cite{XingHu2020DeepCC} uses the same \textbf{Type-4} as Deepcom to obtain the flattened sequence. It also employs the seq2seq model for code summarization and utilizes two GRU as encoders for NCS and flattened sequence, respectively, to acquire both lexical and structural information of code fragments.
GRU has one less gate and relatively fewer parameters than LSTM, so it is easier to converge and less computationally expensive, while having similar results in most tasks, but LSTM performs better with larger data sets, so the choice of LSTM and GRU in code sequence tasks needs to consider both data set size, training effect and training time.

\paragraph{\textbf{Transformer}} 
As previously described,  Transformer is a more complex and powerful model compared to RNN, and more and more work has been done in recent years using transform to learn flattened sequences to accomplish related tasks.
SLM\cite{alon2020structural} uses \textbf{Type-2} to represent the code as a path from the root node and all leaf nodes to the target node. SLM uses LSTM to obtain the representation of all paths separately and then uses Transformer contextualize the path representation of all leaf nodes to the target node, and at the same time adds the position index in the parent node to the path representation from the root node to the target node, and passes the attention mechanism. The final vector is obtained by combining the path representation to make predictions on the target node.
Kim et al.\cite{SeohyunKim2020CodePB} propose three methods based on Transformer to better predict the token that the developer is about to input: 1. pathTrans, uses \textbf{Type-2} to serialize the AST using the path from the leaf node to the root node; 2. TravTrans, uses \textbf{Type-1}, and uses the method of preorder first traversal to obtain sequences from the AST; 3. TravTrans+, uses \textbf{Type-3}, adding a matrix that saves the unique path between two nodes in TravTrans to enhance the Transformer's self Note the block. And the experiment proves that TravTrans+ works better.
Liu et al.\cite{FangLiu2020ASN} uses both \textbf{Type-1} and \textbf{Type-2} to complete the prediction of the code. It uses Transformer-XL to encode the sequence obtained by preorder traversal and uses Bi-LSTM to encode the path from the target node to the root node, preserving the hierarchical information of the target node.
TreeBERT\cite{JiangXue2021TreeBERTAT} employs \textbf{Type-2} to represent the entire code fragment using multiple paths from the root node to the leaf nodes in the AST and then completes the pre-training using the modified Transformer architecture. The position embedding of each node is generated from its hierarchical information in the AST and the position information of its parent node to make better use of the structural information of the code.
Since Transformer requires a lot of data and a lot of computational resources, subsequent work should consider more than just model performance when using Transformer.




\section{Graph-based Models}
\label{graph-models}

While the sequential model with serialized structure as input is simple and visible in many works, the linear order of code snippets is inevitably losing hierarchical syntactic information. Therefore, recent works pay more attention to capturing the syntactic information of the code. Graph-based models for code understanding are described and categorized in this section, based on the structures used in the methods without serialization. 

The structures used in graph-based models are usually AST and Flow Graph, including CFG and DFG, introduced in section \ref{AST} and \ref{flow_graph}. We also introduce models that use other structures in section \ref{other structure}. Unlike the sequence-based models, these structures of a program are considered as a tree structure or graph structure. In this section, we first introduce the transformation conducted in these structures and then categorized different models in AST, flow graphs, the combination of them, and other rarely seen structures generated from source code.

\subsection{Structure Transformation}
\label{graph transformation}

Different from the transformation of sequences, the modifications of graph structures tend to add nodes or edges based on one typical graph structure. We summarize the structure's transformation of the graph into three categories, the first two transformations are proposed on the basis of preserving a graph structure for further learning the representation of the graph, while the last transformation is extracting the information of graph and constructing a new mechanism or DSL(Domain-specific language) without keeping the program graph. 

\paragraph{\textbf{Type-1: Adding Edges}}
In AST-based structure, methods treat the structure from two different perspectives for better using the structural information. One takes the structure as a "tree", which means that the structure has directed edges with the hierarchy preserved. The other extends the AST structure by adding various types of edges and eventually makes the edge bidirectional, which makes the original AST structure a "graph". In two perspectives, adding edges are the most common transformation to preserve more information of structures. Allamanis et al. \cite{Allamanis2018}  use AST as the backbone and add additional edges for capturing data-flow information. The edges contain types derived from AST(e.g. Child and NextToken) and from semantic(e.g. LastUse, LastWrite).  Wang et al. \cite{Wang2021}  extend the AST with parent-child edges. Dinella\cite{dinella2020hoppity} add SuccToken edges between the leaf nodes.

\paragraph{\textbf{Type-2: Combination}}

Some methods use AST-based and Flow-graph-based structures in combination, such as the structure CDFG, which is a combination of CFG and DFG. Other structures for graph-based models have been proposed, as well as some new structures based on the three basic structures. The combination will be discussed and introduced in Section \ref{Combined Structures}. 

\paragraph{\textbf{Type-3: Information Extracting}}
Some methods choose not to preserve the graph structure of AST, Flow graphs, or others. Instead, they propose new structures, mechanisms, or DSL based on the syntactic and semantic information extracted from these graph structures. It is not the priority of our introductions in graph-based models, but still an important kind of structures transformation from graph structures. For example, Raychev et al. \cite{Raychev2016} propose a method to build probabilistic models of code and generate DSL called TGEN, for traversing AST and accumulating a conditioning context. Cvitkovic et al.  \cite{Cvitkovic2019}  propose a Graph-Structured Cache for the out-of-vocabulary problem. An edit DSL called Tocopo\cite{Tarlow2020} is created for code editing in bug fixing problems, which contains token expressions, copy expressions, and pointer expressions. Tocopo is a sequence of edit operations that represents the modification of an AST. 

\subsection{AST-based Structures}

As previously mentioned, the following models are considerably varied as a result of the differences between the perspectives of the AST structure.

\subsubsection{Tree perspective} 
\
\newline \par
Due to the large and deep structure of the tree structure, certain methods tend to use recursion of the tree structure to reduce the complexity of programs, especially passing the information from sub-trees to the full tree. Shi et al.\cite{Shi2021} propose a method to split and reconstruct the whole AST of a program into subtrees hierarchically to get the representation of the code snippets. ASTNN \cite{Zhang2019a} uses a preorder traversal algorithm to split an AST into a set of statement subtrees, recursively encodes them to vectors, and eventually learns the representation of source code through the captured naturalness by BiGRU and RvNN. The method can reduce the difficulty in training. 

Most methods tend to modify the Recursive Neural Networks or Seq2seq model based on the structure of AST, such as AST-based LSTM \cite{Wan2018} and tree-based Seq2seq model \cite{Chakraborty2020}.  The Tree-LSTM is one of the most often modified methods in models of code understanding. Tree-LSTM \cite{Tai2015} is a generalization of the standard LSTM for tree structures that composes the state from an input vector and hidden states of arbitrarily multiple children units. However, when the tree has ordered nodes, such as AST, the original Tree-LSTM is unable to manage the circumstance. To address the problem, a Multi-way Tree-LSTM model is developed, which extends the Tree-LSTM model. The model captures the interaction information between nodes by adding bidirectional LSTM to each gate before linear transformation to encapsulate the information of ordered children. In the code summarization task, the Multi-way Tree-LSTM learns the information in ASTs more effectively than a sequence model.

Convolutional Neural Network is able to train easier and requires less time with the parallel computing mechanism compared to Recursive Neural Networks. Meanwhile, the adaptations to the vanilla CNN, such as tree-based CNN, have demonstrated their efficacy on code comprehensive tasks. Mou et al. \cite{mou2014tbcnn} propose a Tree-Based Convolutional Neural Network (TBCNN) for the program classification task. The convolution layer can capture the information from AST. Chen et al. \cite{Chen2019} propose a tree-based  LSTM  over API-enhanced AST for clone detection. The original AST is modified by adding a new node type to identify the API name. The model called TBCAA learns the representation of code using tree-based CNN, where each convolution kernel has a triangle shape. 

\subsubsection{Graph perspective}
\
\newline \par
To encode the AST structure, most methods consider the structure as a graph and use a graph neural network. Two types of GNN are the most used model in the following graph-based models of AST, which are the Gated Graph Neural Networks(GGNN) and the Convolutional Graph Neural Networks. We also introduce other graph neural networks such as Graph Attention Networks that are used particularly for the AST structures. 

\paragraph{\textbf{Gated Graph Neural Network}} Gated Graph Neural Network, which is a kind of Recurrent Graph Neural Networks, is the first developed graph neural network for code-related tasks when Li et al.\cite{li2015gated} proposed Gated Graph Neural Network to infer formulas for program verification. GGNN updates the state of a node with a GRU cell, with the information from neighboring nodes and the node state in previous timestamp. Fernandes et al. \cite{Fernandes2019} modify GGNN to a framework to extend existing sequence encoders and conduct the experiments on three summarization tasks. Graph-based Grammar Fix (GGF) \cite{Wu2020} use a mixture of GRU and GGNN as an encoder to encode the sub-AST (created by the erroneous code) and a token replacement mechanism as a decoder to generate the code fixing actions. Graph2diff\cite{Tarlow2020} uses an encoder-decoder framework to predict the diff, also described as edit operations. The model takes the source code, bug configures files and compiler diagnostic messages as the graph input. The GGNN is used in the encoder stage to explicit the structure information of code. Although GGNN has shown its ability to learn the syntactic information from the AST of the graph, as a kind of Graph Neural network, it can only aggregate local information rather than global information. Therefore, some works also combine it with sequential models, such as Graph Sandwiches \cite{hellendoorn2019global}. Furthermore, since GGNN employs RNN, it requires more memory to retain the hidden state of all nodes in the graph, despite the fact that it eliminates the need to constrain parameters to ensure convergence.

\paragraph{\textbf{Convolutional Graph Neural Network}} Convolutional Graph Neural Network(convGNN) learns the representations of nodes based on the node vector and the neighbors of the node in the graph. The information from neighbors is combined through the aggregation process.  Compare to Recurrent Graph Neural Networks such as GGNN, Convolutional Graph Neural Networks can stack multiple graph convolutional layers to better propagate the information across nodes, which can combine the information from neighbors. LeClair et al.\cite{LeClair2020} use convolution GNN to encode the AST nodes and edges for code summarization. The ConvGNN layer's input is the AST token embedding and edges. after the ConvGNN layer, the model uses an attention mechanism to learn the important tokens in the source code and AST. Ji et al.\cite{ji2021code} also use GCN to encode AST for the code clone task.  Liu et al.\cite{Liu2021} propose a task of code documentation generation for Jupyter notebooks.  When generating documentation, the model HAConvGNN considers the relevant code cells and code token information. Convolutional GNN layers and a GRU layer are included in the encoder for code cells' AST. The output of the GRU layer will be the input of a hierarchical attention mechanism to better preserve the graph structure. 

\paragraph{\textbf{Other Graph Neural Networks}} There are some methods using other Graph Neural Networks. Different from GCN, Graph Attention Network(GAT) \cite{Velickovic2018} performs self-attention mechanism on message passing stage to learn the graph representation. Wang et al. \cite{Wang2021} models the flattened sequence of a partial AST as an AST graph. To reduce the information loss, the parent-child relation and the positional information are recorded. Further, three types of AST Graph Attention Blocks are proposed to capture the structural information for learning the representation of the graph. Compare to GCN and GGNN, the GAT model improves the explainability of code completion or code summary tasks due to the utilization of the attention mechanism. Hoppity\cite{dinella2020hoppity} uses another type of GNN, which is Graph Isomorphism Network(GIN), as external memory to encode the AST of a buggy program, further using a central controller implemented by LSTM to predict the sequence of actions to fix bugs. The controller will expand or decline the memory when the graph structure is changed. 

\subsection{Flow-Graph-based Structures}

As previously introduced, flow graphs are separated into two types: control flow graphs (CFG) and data flow graphs (DFG). Although a flow graph is more likely to be seen as a graph, there are few works that treat flow graphs as trees. For example, BAST \cite{Lin2021} splits the code of a method according to the blocks in the dominator tree of CFG and generates the corresponding AST for the split code. The split ASTs' representations are used in the pre-trained stage by predicting the next split AST in the dominator tree. The CFG is only used in splitting AST but can make the model more efficient and scalable for large programs. 

The works that consider the CFG from a graph perspective and apply deep learning methods to represent the CFG are described in the following. The attributed Control Flow Graph (ACFG) is a common pre-processing step in some of the following works, especially in binary code similarity detection, such as Genius\cite{Feng2016}, Gemini \cite{Xu2017} and BugGraph \cite{Ji2021a}. There are also some typical modifications for CFG, for example, lazy-binding CFG \cite{Nguyen2018}, inter-procedural CFGs(ICFG) \cite{Duan}. 

\paragraph{\textbf{Convolutional Neural Network}} 
Convolutional neural network (CNN) has translation invariance in many training data, therefore, it can not only capture the semantic information of code but is also suitable for order-aware modeling. Nguyen et al. \cite{Nguyen2018} use CNN on the adjacency matrices converted by the lazy-binding CFG. The CFG will be converted into a pixel image, and later with a CNN model to recognize whether the target object is appears in the image. The method can be applied for malware detection. Yu et al. \cite{Yu2020} use Bert and CNN to learn CFG graph embedding, which can include semantic, structural, and order information. During the adjacent node prediction task, the Bert model is used to pre-train tokens and block embeddings. The order information of CFGs is extracted using CNN models. As indicated previously, CNN models are employed as a component for learning order information or for downstream tasks rather than directly for constructing the representation of the graph of code data.

\paragraph{\textbf{ Convolutional Graph Neural Network}}
DGCNN\cite{zhang2018end}, as one of the ConvGNN, is proposed with a similar pooling strategy SortPooling. The approaches can allow attributed information to be aggregated quickly through neighborhood message passing, therefore, it is suitable for embedding structural information into vectors for further classification. To solve the malware classification challenge, Yan et al.\cite{Yan2019} use DGCNN to embed CFGs. The CFG will first be converted to an attributed CFG, with the code characteristics defining the attributes. The DGCNN is used to aggregate these attributes with an adaptive max pooling to concatenate the layer outputs.

\paragraph{\textbf{ Graph Attention Network}}

Li et al.\cite{Li2020} propose an event-based method CSEM for clone detection. GAT extracts the context information for each statement, which is modeled by the events that are embedded to capture execution semantics. BugGraph\cite{Ji2021a} compares the source-binary code similarity in two steps: source binary canonicalization and code similarity computation. In the code similarity computation step, BugGraph computes the similarity between the target and the comparing binary code. After disassembling both codes, each function will construct its ACFG and use GAT with the triplet loss as the output of the model to generate the embedding of each graph.

\paragraph{\textbf{Other Graph Neural Networks}} As previously mentioned, Yu et al. \cite{Yu2020} also uses  MPNN\cite{Gilmer2017} to compute the graph embedding of a control-flow graph in order-aware modeling. BugLab \cite{Allamanis2021} uses standard message-passing graph neural networks to represent the graph of code entities. The graph includes syntactic entities and relations about the intraprocedural data and control flow, types, and documentation. BugLab trains two models, a selector model and a detector model to predict the rewrite on code snippet. The select model introduces buggy code and the detector model detects and repairs the bugs. Wang et al.\cite{Wang2020GINN} propose a new graph neural architecture called Graph Interval Neural Network(GINN) to learn the code embeddings. GINN takes the program's control flow graph as input and abstracts it using three operators. The CFG is partitioned into a series of intervals using the partitioning operator. Messages between intervals passing are restricted. Then the heightening operator is applied to replace each activate interval with single created nodes until the sufficient propagation point is reached. The lowering operator will finally recover the original CFG. GINN model use only looping construct to learn feature representation, and the method shows improvement across program embeddings.

\subsection{Combined Structures}
\label{Combined Structures}
In some cases, the two types of flow graphs do not appear separately as well as AST. The following part will focus on how these structures are combined and what information can be retrieved from them.

\paragraph{\textbf{Control Flow and Data Flow}} For the combination of two types of graphs, some novel value graph is presented such as the program's Interprocedural Value-Flow Graph (IVFG). The IVFG is created using LLVM-IR, which combines code control-flow and alias-aware data-flows. IVFG is a directed graph with multiple edges. Flow2Vec\cite{Sui2020} is a novel approach for embedding the code with IVFG. Pre-embedding, value-flow reachability via matrix multiplication, and high order proximity approximation are the three steps in the method. Brauckmann et al.\cite{Brauckmann2020} use AST or control-data flow graphs (CDFGs) as the input of the predictive model to learn the code representation. The core of the predictive model is the GGNN in the embedding propagation layer. The model has shown its effectiveness in two complex tasks on OpenCL kernels, the CPU/GPU mapping, and Thread Coarsening. Deepsim \cite{Zhao2018} encodes both code control flow graph and data flow graph into a semantic matrix and uses a deep learning model to measure function similarity. The semantic matrix contains three features: variable features, basic block features, and relationship features between variables and basic blocks.

\paragraph{\textbf{AST and Flow Graphs}} Allamanis et al. \cite{Allamanis2018} propose a method based on GGNN to construct graphs for source code,  naming variables, and detecting variable misuse. The program graph combines both AST and the data flow graph, which contains both syntactic and semantic structure information of source code. Some works combine AST and CFG for multi-modal learning for generating the hybrid representation of code, such as \cite{Wan2019}.  Devign \cite{Zhou2019} utilizes GGNN with a Conv module for graph-level classification. The graph representation is based on AST structure and added multiple types of edges from CFG, DFG, and NCS. The Conv module learns the hierarchy information of representation of node features. Multi-modal learning can reduce the limitation of the approaches only using AST to represent the code. The combination of AST and Flow Graphs can cover extra semantic information of source code.

\subsection{Other Structures}

This section will discuss other structures that differ significantly from the AST and flow graphs. Although the AST and flow-graphs have been employed in most previous works, a program project still contains other graphs. Some are traditional graphs, such as the Program Dependency Graph and UML Diagrams presented in \ref{other structure}, while others are recently proposed structures for better semantic-aware or structure-aware code understanding.

\paragraph{\textbf{Program Dependence Graph}}  Li et al.\cite{Li2019a} propose an attention-based neural network to learn code representation for the bug detection task. The global context is extracted by the Program Dependence Graph and Data Flow Graph using Node2Vec, while the local context is extracted by previous bug fixes and AST paths. The global context and local context will be unified as the path vector for further analysis.

\paragraph{\textbf{UML Diagrams}} CoCoSUM\cite{Wang2021CoCoSum} model the UML diagrams for code summarization task. The framework encodes the class names with a transformer-based model as the intra-class context and the UML diagrams with a Multi-Relational Graph Neural Network (MRGNN) as the inter-class context. The two kinds of embeddings together with the embeddings of the token and AST will be passed to an attention-based decoder to generate code summaries. 

\paragraph{\textbf{Code Property Graph}} Yamaguchi et al.\cite{Yamaguchi2014} model three types of code-related structures: ASTs, CFGs, and Program Dependency Graph \cite{Ferrante1987} as property graphs and combine them into code property graph. The newly proposed data structure enables characterizing the vulnerability type through graph traversals. Liu et al.\cite{Liu2020HybridGNN} use the code property graph and propose a Hybrid GNN framework in the code summarization task. The framework fuse the static graph and dynamic graph to capture the global information of graphs.

\paragraph{\textbf{Program Feedback Graph}} Yasunaga et al.\cite{Yasunaga2020} introduce a program feedback graph to model the reasoning process in program repair task. The nodes in the program feedback graph consist of tokens in the diagnostic arguments, the occurrences in the source code, and the remaining identifiers in the code. The framework DrRepair uses LSTM to encode the source code initially and GAT to further reason over the graph.

\section{Discussion and Comparison}
\label{discussion}
We have introduced two types of models in code understanding: sequence-based and graph-based models in section \ref{seq-models} and Section \ref{graph-models} respectively.

Sequence-based models are models for processing code sequences such as NCS and flatten sequences obtained from AST by several transformation methods introduced in Section \ref{seq-trans}. Traditional statistical language models such as N-gram models are used in a large amount of early works. With the development of deep learning,  word2vec and DBN models have also been used for code representation and downstream tasks. Furthermore, CNN which has revolutionized the computer vision field, can effectively capture rich structural patterns in sequences and are therefore naturally adopted for code sequence tasks. However, the most dominant models among sequence-based models are vanilla RNN models with their variants, such as LSTM, GRU, and Bi-LSTM, which are tailored for sequence modeling tasks. In recent years, transformer-based models that incorporate self-attention blocks have made major contributions to code sequence-based models.

Graph-based models are models that use the graph structures generated by codes such as AST and Flow Graphs to capture structural information. The transformation of graph structures has also been introduced in Section \ref{graph transformation}. Some of these structures, especially AST can be seen as a tree or a graph, which leads to a different way to process the structures. From a tree perspective, these models use RNN models designed for tree structures, such as Tree-LSTM. From the perspective of the graph, CNN is also one of the most used models, while graph neural networks play an important role in the pipeline to learn the representation of nodes or graphs, such as GGNN, GCN, GAT, or MPNN with both AST and flow graphs. Above AST and Flow Graphs, there are also some combination structures and other rarely seen structures, for example, UML diagrams, Program Dependency Graph and so on used in graph-based models.

The two categories of models have both shown the effectiveness on code representation and downstream tasks which we will introduce in Section \ref{tasks}. In this section, we emphasize three differences between the two types of models:

\textbf{First, sequence-based models and graph-based models view data in a different perspective.} Because sequence-based models treat code as a collection of sequences, they must properly capture the relationships between sequences that correspond to semantic and syntactic information in the source code. Graph-based models view code data as a tree or a graph. Since nodes and edges in code graphs are rich in structural information, these models learn the representation in graphs to better understand the code. However, the models that use only sequential data are neglecting the syntactic or semantic information in structures, while the models learn solely on graph structures ignore the sequential information. As previously introduced, the boundary of the two kinds of models is not clear if we divide them with the information they used. There are sequences flattened from the AST structure that automatically combine the structural information in AST. In Graph-based models, some methods also use sequential information in code.

\textbf{Second, sequence-based and graph-based models are both use serial models but with distinct proposes and scenarios.} RNN and Transformer are the most commonly used serial models in deep learning models in code understanding, but the two types of models use RNN and Transformer with different proposes. Sequence-based models use RNN and Transformer as the core component of the models to learn relationships in code sequences. The RNN\cite{Veselin2014CodeCW, Tal2018NeuralCC} is used in the sequence-based models to tackle the problem of information passing in the code sequence. LSTM\cite{dam2016deep,iyer2016summarizing} and GRU\cite{AlexanderLeClair2019ANM, XingHu2020DeepCC} which are the variant of RNN use the gating mechanism to better transfer useful information and solve the problem of gradient explosion and disappearance in code sequences. Transformer-based models\cite{ahmad2020transformer,alon2020structural} are more commonly employed to address the issue of global reliance and use positional embedding to maintain the order information of code sequence, allowing for better learning of the entire code information.

Although RNN-based model designed for serial data, they are widely used in graph-based models as sequences are the original form of code that contains the context information. Furthermore, although Graph Neural Network can capture the structural information of structures, the context information cannot be preserved once the source code is converted to a graph structure. Therefore, approaches in graph-based models will also use RNN such as LSTM or GRU to gather long-distance information. Some works use RNN for tree structures such as RvNN\cite{Shi2021}, or tree-LSTM\cite{Wan2019}. Most of the works combined the RNN with GNN as the whole model frameworks, mainly categorized as the following three usages: 1) Conducting bidirectional GRU\cite{Zhang2019a} or bidirectional LSTM as the encoder of the models, 2) using GRU to fuse resulting vectors after GNN components\cite{Liu2020HybridGNN}, 3) using RNN such as LSTM as decoder\cite{Wang2020GINN}. Graph Sandwich Structure\cite{hellendoorn2019global} is one of the most typical works that combine GNN with RNN and Transformer structure to combine both local and global information.

\textbf{Third, Attention mechanism are used in sequence-based and graph-based models but with various modifications.}    

The attention mechanism is widely used in neural machine translation, especially in encoder-decoder frameworks. The attention mechanism can learn which words are important and further make predictions according to the importance of words. In code representation-related works, attention mechanisms are conducted in both sequence-based models and graph-based models.

In sequence-based models, the attention mechanism is primarily used to pay attention to the relationships between tokens in code sequences, including the relationships between input sequence tokens, between output sequence tokens and input sequence tokens, or between output sequence tokens, in order to facilitate more efficient information transmission. The Transformer-based models use self-attention to focus on the three relationships mentioned above while the other models pay more attention to the relationship between output sequence tokens and input sequence tokens. Attention mechanisms are added to RNN-based models by CodeNN\cite{iyer2016summarizing} and other models, allowing the model to make greater use of crucial token information while creating code summary or predicting the next token.

Some work in graph-based models conduct attention mechanism before output layer and create a context vector to predict the next token in the sequence, such as LeClair et al.\cite{LeClair2020}. Others make modifications on vanilla deep learning models with attention mechanism, for example, using attention mechanism in message propagation process of GCN to hierarchically update the embeddings\cite{ji2021code}, combining attention mechanism in GRU layer and Convolutional layer to encode the order of the nodes in a path \cite{Li2019a}. Different from sequence-based models, some works modify the original attention mechanism to better suit the hierarchical structure. Liu et al\cite{Liu2021} propose low-level attention and high-level attention. The former attention module attends the token in a sequence while the latter attends the code cell in the AST tree.


\section{Tasks}
\label{tasks}
\par We categorized the tasks used in code understanding into the following downstream tasks and summarized the sequence-based models and graph-based models on each downstream tasks.
\subsection{Code Generation}
Code Generation can provide varying levels of granularity of code output depending on the input. Code prediction in IDE that anticipates blank areas of input code snippets, such as method name prediction, next token prediction, and so on, is a typical code generation task. It's also a kind of code generation task to generate snippets of code from natural language. Code generation considerably enhances developers' efficiency and has been extensively researched in both industry and academia. The relevant approaches are summarized in Table \ref{Code_Generation_table}, which is based mostly on the structure used by the model in the code generation task.

\subsubsection{Method name generation}
This task generates a summary method name based on the given method code body, which can help the code be more understandable, maintainable, and invocable. The formalization of the method name generation task is as follows:
\par Given a code body of a method with n tokens $S={t_1,t_2,...t_n}$, the generative model can output the method name.

\subsubsection{Next token generation}
This task predicts the tokens including API calls that will be inputted by the developer. The models of the next token generation typically provide a list of tokens in order of probability depending on the developer's previous input. It can substantially speed up development. The formalization of the next token generation task is as follows:
\par Given a partial code snippet with $n-1$ tokens $S={t_1,t_2,...t_{n-1}}$, the generative model can generate a token list of $t_n$ sorted by probability.

\subsubsection{Expression generation}
Compared to the next token generation, this task is a more granular code prediction task based on existing code fragments. It generates complete code expressions with certain functions such as conditional statements, loops, etc. It also includes the task of predicting missing code statements. The formalization of the next expression generation task is as follows:
\par Given a partial code snippet with $n$ tokens $S={t_1,t_2,...,t_{n-1}}$, the generative model can generate a whole code expression $S_m={t_{n},t_{n+1}...,t_{n+m-1}}$ with a specific function. Or given a code with a missing piece $S={t_1,t_2,...,missing piece,...,t_{n}}$, the generative model generates the missing piece code.  
 
\subsubsection{Function generation}
This task can generate corresponding code snippets based on the user's description of the code's functionality, which can be seen as a natural language to the code translation process. Excellent natural language to code translation techniques can enable non-specialists to develop accordingly, but work in this area is still immature and needs to be further developed. The formalization of the function generation task is as follows:
\par Given a natural language $D$ for the functional description of the code $S$, the generative model can generate the $S={t_1,t_2,...,t_{n}}$ according to $D$.

\begin{table}
\centering
\captionsetup{}
\caption{Code Generation}
\label{Code_Generation_table}
\begin{tabular}{>{\centering\hspace{0pt}}m{0.088\linewidth}|>{\centering\hspace{0pt}}m{0.13\linewidth}|>{\centering\hspace{0pt}}m{0.15\linewidth}|>{\centering\hspace{0pt}}m{0.12\linewidth}|>{\hspace{0pt}}m{0.48\linewidth}} 
\hline\hline
Type                                                            & Reference    & Model                       & Generation     & \multicolumn{1}{>{\centering\arraybackslash\hspace{0pt}}m{0.48\linewidth}}{\textcolor[rgb]{0.2,0.2,0.2}{Description}}  \\ 
\hline
\multirow{10}{*}{\Centering{}Sequence} &  Code2seq\cite{Alon2018code2seq} & BiLSTM  & method name    & Represents code by a collection of paths between terminal nodes n the AST                                             \\ 
\cline{2-5}
                                                                & Code2vec\cite{Alon2019code2vec} & LSTM & method name    & Generates code's representation by AST path                                                                            \\ 
\cline{2-5}
                                                                & SLM\cite{alon2020structural}      & LSTM          & expression & Generates missing code using all the paths to it                                                                       \\ 
\cline{2-5}
                                                                & Pythia\cite{Alexey2019Pythia}   & LSTM                        & next token      & Predicts the methods and API calls by flattened AST                                                                    \\ 
\cline{2-5}
                                                                & \cite{ChangLiu2017NeuralCC}       & LSTM                        & next token     & Uses the sequence obtained by traversing the AST to predict the next possible node                                     \\ 
\cline{2-5}
                                                                &\cite{2017pointerMixtureNetork}           & LSTM, Pointer Network       & next token     & Generates code through LSTM or pointer network                                                                        \\ 
\cline{2-5}
                                                                &\cite{SeohyunKim2020CodePB}          & Transformer                 & next token     & Feeds different paths to Transformer                                                                                   \\
\cline{2-5}
                                                                &\cite{FangLiu2020ASN}          & Transformer-XL                 & next token     & Feeds different paths to Transformer-XL with multi-task learning                                                                                 \\ 
\cline{2-5}
                                                                &\cite{UriAlon2018AGP}          & Word2vec                 & method name     & Feeds different paths with up/down momentum to Word2vec                                                                                \\ 
\cline{2-5}
                                                                &API2Vec\cite{TrongDucNguyen2017API2vec}          & Word2vec                 & next token     & Feeds different API paths to Word2vec                                                                                 \\                                
\hline
\multirow{4}{*}{\Centering{}Graph}   
   & DEEP3 \cite{Raychev2016} &       Decision tree      &  next token     &  Represents code in a DSL called TGEN and build probabilistic models with decision trees\\
\cline{2-5}
   & \cite{Brockschmidt2019}    &       GRU,GGNN    &      expression &  A generated model on ExprGen task which first obtains a graph by attribute grammars and later compute the attribute representations with GGNN and GRU  \\

\cline{2-5}
   &  CCAG\cite{Wang2021} &       GAT                &    next token   &  Uses AST Graph Attention Block(ASTGab)  to model the flattened sequence of AST to capture different dependencies\\
\cline{2-5}
   &  \cite{Allamanis2018} &       GGNN     &  method name     &Constructs graphs by adding different types of edges and use GGNN to learn the representation of the graph  \\
\cline{2-5}
   &  Graph-Structured Cache\cite{Cvitkovic2019} &    MPNN,CharCNN    &   method name    &  Introduces a Graph-Structured Cache representing vocabulary words as additional nodes, uses MPNN and CharCNN to generate outputs to further address open vocabulary issue\\
 
\hline\hline
\end{tabular}
\end{table}

\subsection{Code Summarization}
Code summarization is the work of using natural language to provide a concise explanation of the code's functioning. It can help enhance the code's readability, as well as the developer's efficiency in understanding the program. The code summarization task can be thought of as a translation of the input code into natural language, hence a seq2seq model architecture is commonly used. In the encoder phase, \cite{Hu2018DeepCC,XingHu2020DeepCC,AlexanderLeClair2019ANM} convert the input NSC or Flattened sequence into a context vector, and then in the decoder phase, they construct the words in the summarization one by one based on the context vector. To increase the effectiveness of code summarization, techniques such as the attention mechanism are applied to the task. The relevant techniques are summarized in Table \ref{Code_Summarization_table} based on the model's relevant code structures.

\begin{table}
\centering
\captionsetup{}
\caption{Code Summarization}
\label{Code_Summarization_table}
\begin{tabular}{>{\centering\hspace{0pt}}m{0.104\linewidth}|>{\centering\hspace{0pt}}m{0.104\linewidth}|>{\centering\hspace{0pt}}m{0.089\linewidth}|>{\hspace{0pt}}m{0.642\linewidth}} 
\hline\hline
Type                                                            & Reference           & Model & \multicolumn{1}{>{\centering\arraybackslash\hspace{0pt}}m{0.631\linewidth}}{\textcolor[rgb]{0.2,0.2,0.2}{Description}}  \\ 
\hline
\multirow{7}{*}{\Centering{}Sequence} & DeepCom \cite{Hu2018DeepCC}        & LSTM  & Employs the seq2seq model which uses LSTM as encoder and decoder to generate code fragment summaries                                                                                            \\ 
\cline{2-4}
                                                                & Hybrid-Deepcom \cite{XingHu2020DeepCC} & LSTM  &Utilizes two GRU as encoders for NCS and flattened sequence                                             \\ 
\cline{2-4}
                                                                & Ast-attendgru \cite{AlexanderLeClair2019ANM}  & GRU   & Uses two GRU with attention mechanism to process NCS and flattened sequence respectively for getting context vector                                         \\ 
\cline{2-4}
                                                                &\cite{Allamanis2016}                &CNN       & Uses Convolutional Attention Network to summarize the code                                                                        \\ 
\cline{2-4}
                                                                &\cite{iyer2016summarizing}                &LSTM      & Uses LSTM with attention to produce sentences that describe code snippets                                                                        \\ 
\cline{2-4}
                                                                &\cite{JianZhang2020RetrievalbasedNS}                &search engine      & Retrieves flattened sequence on the search engine to obtain syntactically similar code fragments                                                                         \\ 
\cline{2-4}
                                                                &\cite{ahmad2020transformer}                &Transformer      & Employs transformer for code summarization
                                                                \\
\hline
\multirow{9}{*}{\Centering{}Graph}    & \cite{Wan2018}        &   RNN, Tree-RNN    & Uses a attention layer to fuse two representations, one from the structure of source code with AST-based LSTM, the other from the sequence with LSTM \\ 

\cline{2-4}
  & \cite{Fernandes2019}  &  GGNN     &   Uses sequential encoder and a GGNN to generate the representations      \\
  
\cline{2-4}
  & TBCAA\cite{Chen2019}  &    Tree-based LSTM   &    Uses Tree-based convolution for API-enhanced AST      \\
\cline{2-4}
  & \cite{LeClair2020}  &   ConvGNN    &   Encodes the node token embedding with recurrent layer and a ConvGNN, later uses an attention layer to learn important tokens       \\
\cline{2-4}
  & Flow2Vec \cite{Sui2020}  &   Flow2Vec    &   Pre-embeds the interprocedural value-flow graph, considers the reachability via matrix multiplication problem and uses it to approximate the high-order proximity embedding       \\

\cline{2-4}
  & BASTS \cite{Lin2021}  &   Tree-LSTM    &   Uses Tree-LSTM and Transformer architecture to combine the representations of split AST and source code       \\
\cline{2-4}
  & CoCoSum \cite{Wang2021CoCoSum}  &    Transformer, Multi-Relational GNN   &    The global encoder contains an MRGNN to embed the UML class diagrams and a  transformer-based model for embedding the class names, while the local encoder uses the GRU      \\

\cline{2-4}
  & HybridGNN \cite{Liu2020HybridGNN}  &   GNN    &   A hybrid message passing GNN based which fuse the static and dynamic graph       \\

\cline{2-4}
& \cite{Shi2021} &   RNN, attention    & A convolutional attention network for extreme summarization of source code \\ 
\cline{2-4}
  & \cite{Liu2021}  &   HAConvGNN    &  Hierarchically splits the AST into subtrees, learns the representation of split AST, and reconstructs them to combine the structural and semantic information with RvNN        \\
 
\hline\hline
\end{tabular}
\end{table}
\subsection{Code Search}
Code search is a information retrieval task. The input of code search is the query which is natural language description or a code snippet. And  the output is the best matching code snippets for input. The goal of Code Search is to retrieve code snippets from a large code corpus that most closely match a developer’s intent\cite{JosPabloCambronero2019WhenDL}. Being able to explore and reuse existing code that is match to the developer’s intent is a fundamental productivity tool. Some online sites such as Stack Overflow are popular because of the convenience that searching for code relevant to a user’s question expressed in natural language. 

Some articles refer to this task as \textit{semantic code search} or \textit{code recommendation}, these names emphasize characteristics of code search. It is based on the semantics of the input, and the semantic alignment between input and code snippets is crucial. And the information retrieval nature of this task is that all outputs are retrieved from the code corpus as they originally are. The retrieval nature distinguishes code search and code generation, where the generation task is aiming to synthesize and write codes not only already in the code corpus.

The code search models are summarized in Table \ref{Code_Search_table}.

\begin{table}
\centering
\caption{Code Search}
\label{Code_Search_table}
\begin{tabular}{>{\centering\hspace{0pt}}m{0.104\linewidth}|>{\centering\hspace{0pt}}m{0.104\linewidth}|>{\centering\hspace{0pt}}m{0.15\linewidth}|>{\hspace{0pt}}m{0.662\linewidth}} 
\hline\hline
\par{}Type    & Reference          & Model                                                 & \multicolumn{1}{>{\centering\arraybackslash\hspace{0pt}}m{0.662\linewidth}}{\textcolor[rgb]{0.2,0.2,0.2}{Description}}  \\ 
\hline
\multirow{4}{*}{\Centering{}Sequence} & \cite{Sachdev2018RetrievalOS}                                              & word2vec                                                & Combines natural language techniques and information retrieval methods                                               \\ 
\cline{2-4}
                                                                &CODEnn~\cite{Gu2018DeepCS}                                                       &RNN                                                       &Jointly embeds code snippets and natural language descriptions into a high-dimensional vector space                                                                                                                         \\ 
\cline{2-4}
                                                                &UNIF\cite{JosPabloCambronero2019WhenDL}                                                       &word2vec                                                       &Uses attention to combine per-token embeddings and produce the code sentence embedding                                                                                                                       \\                                                             
\cline{2-4}
                                                                &CARLCS-CNN~\cite{Shuai2020ImprovingCS}                                                      &CNN                                                       &Uses CNN to embed code and query respectively                                                                                                                         \\                 
\hline
\multirow{6}{*}{\Centering{}Graph} & TBCAA\cite{Chen2019}&  tree-based LSTM   &  Tree-based convolution for API-enhanced AST \\ 
\cline{2-4}
& MMAN\cite{Wan2019}&  GGNN, Tree-LSTM   &  A multi-modal Attention Network that uses attention mechanism to capture the information from an LSTM for embedding sequential tokens, a Tree-LSTM for embedding  AST, and a GGNN for representing CFG \\ 
\hline\hline
\end{tabular}
\end{table}

\subsection{Clone Detection}
Clone detection task indicates that there are two or multiple similar code snippets in the same software or system, which is also known as code clones. The code clones can support the modifications by the developers for better reusing them. Code clones can be described as the following 4 types \cite{clone2007}:

\textbf{Type-1} clones are identical code fragments, which may contain slight differences in white-space, layouts, or comments.

\textbf{Type-2} clones are identical code fragments, which may contain the differences of variable names, constants, function names, identifiers, literals, types, layouts, white-space, or comments.

\textbf{Type-3} clones are syntactically similar code fragments with added, deleted, or modified statements.

\textbf{Type-4} clones are semantic similar code fragments that may use different lexical and syntax to express the equivalent semantic.

As the similarities decrease in the four types, the difficulty of detecting the clones increases. The approaches to solve the clone detection is shown in Table \ref{clone_detection_table}. 



\begin{table}
\centering
\captionsetup{}
\caption{Clone Detection}
\label{clone_detection_table}
\begin{tabular}{>{\centering\hspace{0pt}}m{0.104\linewidth}|>{\centering\hspace{0pt}}m{0.104\linewidth}|>{\centering\hspace{0pt}}m{0.089\linewidth}|>{\hspace{0pt}}m{0.642\linewidth}} 
\hline\hline
Type                                    & Reference           & Model & \multicolumn{1}{>{\centering\arraybackslash\hspace{0pt}}m{0.631\linewidth}}{\textcolor[rgb]{0.2,0.2,0.2}{Description}}  \\ 
\hline
\multirow{2}{*}{\Centering{}Sequence} & CCLearner \cite{li2017cclearner}      & DNN    &Extracts tokens from known method-level code clones and non-clones to train a classifier                                                       \\ 
\cline{2-4}  & \cite{white2016deep} & RNN  &Uses RNN modeling sequences of terms in a source code corpus                                         \\ 
\hline
\multirow{3}{*}{\Centering{}Graph}   &   DeepSim \cite{Zhao2018}    &   Feed-forward neural network    &  An approach for measuring code functional similarity that uses two flow graphs as the basis and encodes them with Feed-forward neural network \\
\cline{2-4} &   ASTNN \cite{Zhang2019a}     &   bidirectional RNN    &    Encodes the statement subtree and uses Bi-GRU to model the naturalness of the statements  \\ 
\cline{2-4} & TBCAA  \cite{Chen2019}     &    tree-based LSTM   &  Tree-based convolution for API-enhanced AST\\
\cline{2-4} & DEEPBINDIFF \cite{Duan}      &   Text-associated DeepWalk    & Learns basic block embeddings with Text-associated DeepWalk algorithm, and match them with the k-hop greedy matching algorithm  \\
\cline{2-4} &   \cite{Yu2020}     &  MPNN, CNN     &  Use BERT to pre-train token and block embeddings  on an MLM task, and fine-tune them on 2 graph-level tasks with MPNN, GRU, and CNN\\
\cline{2-4} & CSEM \cite{Li2020}      &   Transformer, GAT, CNN    &  Converts  source code to intermediate representation, generates Node vector matrix and inputs it into GAT later and CNN layer to obtain embedding of code fragment \\
\cline{2-4} & OSCAR \cite{Peng2021}      &    Transformer   &  A hierarchical multi-layer Transformer pre-trained model with a novel positional encoding, contrastive learning with optimization techniques \\
\cline{2-4} & HAG  \cite{ji2021code}      &    GCN   &   Uses GCN with layer-wise propagation and attention mechanism\\
\hline\hline
\end{tabular}
\end{table}

\subsection{Safety Analysis}
Humans are relying more and more on programs and codes to handle various problems in life as computer technology advances, and defects and vulnerabilities in codes can result in significant losses. As a result, it is vital to examine the code's reliability and security. Defects in code can cause programs to fail to run properly, and vulnerability can pose a potential threat to the safe operation of computer systems. Furthermore, Malware is Malicious software which is designed to attack the device. Therefore, in Safety analysis, we categorized three safety-related tasks: defect prediction, vulnerability prediction, and malware classification and further summarize the models for these three categories in Table \ref{Safety_analysis_table}.
 
\subsubsection{Defect Prediction}
Defect prediction helps developers test more effectively while lowering the cost of software development by predicting areas of problematic code.
The two types of defect prediction tasks now accessible are within-project defect prediction (WPDP), in which the training and test sets are from the same project, and cross-project defect prediction (CPDP), in which the test set is different from the training set. 
The formalization of the Defect prediction task is as follows:
\par Given a code snippet with $n$ tokens $S=\{t_1,t_2,...t_n\}$, the predictive model can output a label $y$ which means the code snippet with defects(Buggy) or without defects(clean).
\subsubsection{Vulnerability Detection}
The Vulnerability detection task, including vulnerability detection based on code similarity and code patterns, can prevent code from being attacked and improve the security of code. Some approaches use deep learning for vulnerability detection ,for example, VulDeePecker \cite{Li2018}. The formalization of the Vulnerability detection task is as follows:
\par Given a code snippet with $n$ tokens $S=\{t_1,t_2,...t_n\}$, the predictive model can output a label $y$ which means the code snippet with or without vulnerability.

\subsubsection{Malware Classification}
 Malware classification is one kind of malware detection, which is a binary classification problem. Malware classification can be formalized as: Given a program P, classify P as a normal program or malware.



\begin{table}
\centering
\caption{Safety analysis}
\label{Safety_analysis_table}
\begin{tabular}{>{\centering\hspace{0pt}}m{0.094\linewidth}|>{\centering\hspace{0pt}}m{0.16\linewidth}|>{\centering\hspace{0pt}}m{0.07\linewidth}|>{\hspace{0pt}}m{0.067\linewidth}|>{\hspace{0pt}}m{0.57\linewidth}} 
\hline\hline
Type                                                            & Reference                                                 & Model                                                 & Kind   & \multicolumn{1}{>{\centering\arraybackslash\hspace{0pt}}m{0.57\linewidth}}{\textcolor[rgb]{0.2,0.2,0.2}{Description}}  \\ 
\hline
\multirow{4}{*}{\hspace{0pt}\Centering{}Sequence} & \cite{SongWang2016AutomaticallyLS}                                              & DBN                                                  & Defect & Uses DBN to automatically learn the semantic expression of code                                                 \\ 
\cline{2-5}
                                                                & Seml\cite{liang2019seml}                                       & LSTM                                                  & Defect & Predicts software defect with the help with LSTM                                                        \\ 
\cline{2-5}
                                                                & DeepCPDP\cite{DeyuChen2019DeepCPDPDL}                                        & Bi-LSTM                                                   & Defect & Proposes SimAST and SimAST2Vec                                                  \\ 
\cline{2-5}
                                                                & \cite{SongWang2020DeepSF}                                        & DBN                                                   & Defect & Utilizes the DBN to learn the higher-level semantic characteristics of code AST token                                                  \\

\cline{2-5} &\cite{QuanLe2018DeepLA}  & CNN-BiLSTM  & Malware  & Proposes an approach to enable malware classification by malware analysis non-experts    \\  
\cline{2-5}
&   \cite{DanielGibert2019UsingCN}  & CNN  &Malware  &Proposes a file agnostic deep learning approach for malware categorization to efficiently group malicious software into families based on a set of discriminant patterns extracted from their visualization as images    \\ 

\hline
\multirow{3}{*}{\hspace{0pt}\Centering{}Graph} &   \cite{Allamanis2018}     &   GGNN     &   Defect     & Constructs graphs by adding different types of edges and uses GGNN to learn the representation of the graph \\ 
\cline{2-5}  & MAGIC \cite{Yan2019}     &  DGCNN      &     Defect    &  Extends the standard DGCNN on Weighted Vertices layer and Adaptive Max Pooling to aggregate attributes from graph structures \\ 
\cline{2-5}  &  \cite{Li2019a}     &   GRU, CNN, Attention mechanism     &     Defect    &   Extracts the global (from PDG and DFG) and local (from AST) context with Attention-Based GRU, CNN. \\ 
\cline{2-5}  &  Devign \cite{Zhou2019}     &    GGNN, GRU, CNN    &  Vulnerability      &  Encodes the code into a joint graph structure and uses GGNN with the Conv module to learn the embedding. \\ 
\cline{2-5}  &  BugGraph \cite{Ji2021a}     &   GTN     &     Vulnerability   &  A Graph Triplet-loss Network on the attributed CFG to learn similarity ranking. \\
\cline{2-5}  &  \cite{Brauckmann2020}     &   GGNN     & Malware  &  Uses GGNN for learning predictive compiler tasks on AST and CDFGs           

\\ 
\hline\hline
\end{tabular}
\end{table}

\subsection{Bug Localization}
\label{bug_localization}

Bug localization is a task that localizes the position of bugs in a buggy code snippet. The bug localization can also be seen as the previous step of program repair. Bugs fall into two categories based on how they are discovered: static and dynamic. The static bug location is determined by the control and data dependencies, whereas the dynamic bug location is determined by the program execution. We provide works of bug localization task in Table \ref{Bug_Localization_table}.
The formalization of bug localization task is as follows:
\par Given a code snippet with $n$ tokens $S=\{t_1,t_2,...t_n\}$, the predictive model will predict the position $\lambda$ of buggy token $t_\lambda$ such as the misused variables or operators. The place where the correct token is needed to predict is also called a slot.

\begin{table}
\centering
\caption{Bug Localization}
\label{Bug_Localization_table}
\begin{tabular}{>{\centering\hspace{0pt}}m{0.104\linewidth}|>{\centering\hspace{0pt}}m{0.104\linewidth}|>{\centering\hspace{0pt}}m{0.069\linewidth}|>{\hspace{0pt}}m{0.662\linewidth}} 
\hline\hline
\par{}Type    & Reference          & Model                                                 & \multicolumn{1}{>{\centering\arraybackslash\hspace{0pt}}m{0.662\linewidth}}{\textcolor[rgb]{0.2,0.2,0.2}{Description}}  \\ 
\hline
\multirow{2}{*}{\Centering{}Sequence} & \cite{MarkoVasic2019NeuralPR}                                               & LSTM                                                  & Presents  multi-headed pointer networks for training a model that jointly and directly localizes and repairs variable-misuse bugs                                               \\ 
\cline{2-4}
                                                                &BULNER\cite{JacsonRodriguesBarbosa2019BULNERBL}                                                       &Word2vec                                                       & Proposes a method for Bug Localization with word embeddings and Network Regularization                                                                                                                         \\ 
\hline
\multirow{6}{*}{\Centering{}Graph} &  GGF \cite{Wu2020}        &     GGNN    &  Uses GRU and GGNN as encoder and a token replacement mechanism as decoder to encode  the token information and generate the fixing actions    \\ 
\cline{2-4}  &   \cite{dinella2020hoppity}        &   GAT, LSTM      &    Introduces a program feedback graph and apply GNN to model the reasoning process.  \\ 
\cline{2-4}  &  GINN \cite{Wang2020GINN}        &    GINN     &   Proposes Graph Interval Neural Network, which includes the heightening and lowering operator to learn the representation of CFG    \\ 
\hline\hline
\end{tabular}
\end{table}

\subsection{Program Repair}

Program Repair, also known as bug fix, code refinement, aims at fixing the localized bugs. Some works perform the bug localization and program repair tasks jointly, for example, BugLab \cite{Allamanis2021}. After detecting the bugs, the repair is conducted on the typical line of the program.  Some works combine bug detection and program repair, which predict the location and fix action at the same time with a sequence combining two of them\cite{MarkoVasic2019NeuralPR}.  Other repair tasks will only fix bugs assuming the bug locations already exist, such as CODIT\cite{Chakraborty2020}. 

One of the most popular tasks in automated program repair is the VARMISUSE task proposed by Allamanis\cite{Allamanis2018}. The VARMISUSE task is to automatically detect the variable misuse mistakes in source code and repair it with the correct variable.  In other words, program repair task can be seen as a kind of code generation in the slot where the mistake is detected. However, due to the difference of input and the proposal of generating the code, we consider program repair as a new task and summarize the models in Table \ref{Program_Repair_table}.

Because of the diverse datasets, languages, and granularities of the output, different works have distinct definitions of program repair. For example, the output may be a single word to repair a misused variable or a full sentence to repair a new code snippet line. The formalization of the repair of a token is similar to the formalization in \ref{bug_localization}. After detecting the positions, the program repair task will generate the correct token $t_\lambda'$. Another of the formalizations of the program repair task is as follows:

Given a broken code snippet $S=\{l_1,l_2,...l_k\}$ (with $k$ line) the diagnostic feedback from compiler(the feedback usually contains line number and error message) $I$ , the program repair task is to localize the erroneous line index $n$ and generate a repaired code version $l_n^{'}$ replacing the wrong code of $l_n$. 

\begin{table}
\centering
\caption{Program Repair}
\label{Program_Repair_table}
\begin{tabular}{>{\centering\hspace{0pt}}m{0.104\linewidth}|>{\centering\hspace{0pt}}m{0.104\linewidth}|>{\centering\hspace{0pt}}m{0.069\linewidth}|>{\hspace{0pt}}m{0.662\linewidth}} 
\hline\hline
\par{}Type    & Reference          & Model                                                 & \multicolumn{1}{>{\centering\arraybackslash\hspace{0pt}}m{0.662\linewidth}}{\textcolor[rgb]{0.2,0.2,0.2}{Description}}  \\ 
\hline
\multirow{2}{*}{\Centering{}Sequence} & DeepFix\cite{RahulGupta2017DeepFixFC}                                               & GRU                                                  &Fixes multiple errors by iteratively invoking a trained neural network                                             \\ 
\cline{2-4}
                                                                &TFix\cite{BerkayBerabi2021TFixLT}                                                       &Transformer                                                       &Uses T5\cite{ColinRaffel2019ExploringTL} to accurately synthesize fixes to a wide range of errors                                                                                                                         \\ 
\hline
\multirow{6}{*}{\Centering{}Graph} &  CODIT \cite{Chakraborty2020}        &   LSTM      &     An encoder-decoder model which first learns the structural changes in AST modifications with tree-to-tree model, then predicts the token conditioned on the AST \\ 
\cline{2-4}  &  GGF \cite{Wu2020}        &     GGNN    &  Uses GRU and GGNN as encoder and a token replacement mechanism as decoder to encode the token information and generate the fixing actions    \\ 
\cline{2-4}  &  Hoppity \cite{dinella2020hoppity}        &   GAT, LSTM      &    Introduces a program feedback graph and applies GNN to model the reasoning process  \\ 
\cline{2-4}  &  GINN \cite{Wang2020GINN}        &    GINN     &   Proposes Graph Interval Neural Network, which includes the heightening and lowering operator to learn the representation of CFG    \\ 
\hline\hline
\end{tabular}
\end{table}

\subsection{Pre-training Task}

Pre-training is the process of training a model on a large amount of pre-training data to extract as many features as possible from the data so that the model can better solve the aiming task after fine-tuning in specific dataset. With the proliferation of open source corpora, pre-trained models have emerged for large amounts of code data, such as CodeBERT~\cite{Feng2020CodeBERTAP},CuBert\cite{Kanade2020}, GPT-C\cite{svyatkovskiy2020intellicode},  CodeT5\cite{YueWang2021CodeT5IU}and GraphCodeBERT\cite{guo2020graphcodebert}, which can capture semantic information in code and be quickly and effectively applied to various downstream tasks. The pre-training models are becoming mainstream models for code-related tasks. A series of pre-training models have been proposed by large companies such as Microsoft and Facebook. Some models such as Alphacode\footnote{https://alphacode.deepmind.com/} and Codex\footnote{https://openai.com/blog/openai-codex/} have been applied in practice.

\section{Metrics and Datasets}
\label{metrics_and_datasets}
We describe the metrics and datasets associated with code-related work in this section to serve as a reference for future work.

\subsection{Metrics}
The relevant metrics in the code representation field are divided into two categories: NLP-related metrics and information retrieval-related metrics.

\subsubsection{NLP-related Metrics}
Due to the similarity between code and natural language, many metrics employed in the field of code are generated from the natural language field, particularly the generation tasks such as code summary.
\begin{itemize}
    \item \textbf{Accuracy, Precision, Recall and F1-score} One of the most intuitive performance metrics is accuracy, which is defined as the ratio of correct predictions to total predictions. Precision refers to the percentage of correct predictions of true samples among all predictions predicted as true samples whereas Recall refers to the percentage of correctly predictions of true samples among all true samples. The weighted average of Precision and Recall is the F1 Score. As a result, this score considers both false positives and false negatives. It is more useful, especially when the distribution of classes is uneven. The three metrics are commonly used in classification or prediction tasks.
    \item \textbf{BLEU} BiLingual Evaluation Understudy(BLEU)\cite{Papineni2002BleuAM} is designed for automated evaluation of statistical machine translation and can be used to measure the performance of code summarization and generation tasks. The score is computed as Equation \ref{con:BP} and \ref{con:BLEU}, where the former is the Brevity Penalty(BP) with the length of the candidate translation $c$ and the effective reference sequence length $r$. TBP $p_n$ is the ratio of length n subsequences in the candidate that is also in the reference. And the $N$ in Equation \ref{con:BLEU} is usually set to 4(BLEU-4)\cite{Hu2018DeepCC}.
    
\begin{equation}
\mathrm{BP}=\left\{\begin{array}{ll}
1 & \text { if } c>r \\
e^{(1-r / c)} & \text { if } c \leq r
\end{array}\right.\label{con:BP}
\end{equation}

\begin{equation}
\mathrm{BLEU}=\mathrm{BP} \cdot \exp \left(\sum_{n=1}^{N} w_{n} \log p_{n}\right)\label{con:BLEU}
\end{equation}
    
\item \textbf{Perplexity (PPL)} Perplexity is a great probabilistic measure used to evaluate exactly how confused the NLP models are. It’s typically used to evaluate language models, as well as the dialog generation tasks. PPL is defined as Equation \ref{con:PPL},where $x_i$ is the truth label and  $P(x_i)$ is the model output. A model with lower perplexity assigns higher probabilities to the true tokens and is expected to perform better.
    
\begin{equation}    P P L=\exp \left(-\sum_{i}^{T} P\left(x_{i}\right) \log P\left(x_{i}\right)\right), \forall i \in 0 \ldots T.\label{con:PPL}
\end{equation}

\item \textbf{ROUGE} As opposed to the BLEU score, the Recall-Oriented Understudy for Gisting Evaluation (ROUGE) evaluation metric\cite{ChinYewLin2004ROUGEAP}  measures the recall.It is commonly used in machine translation tasks to assess the quality of generated text. However, because it assesses recall, it is mostly employed in summarization tasks, where evaluating the amount of words the model can recall is more significant. \cite{ChinYewLin2004ROUGEAP} proposes four Rouge methods: 1) ROUGE-N: calculate the recall rate on n-gram, 2) ROUGE-L: consider the longest common subsequence between the generated sequence $C$ and the target sequence $S$, 3) ROUGE-W: improve ROUGE-L and calculate the longest common subsequence by weighting method, and 4) calculate the recall rate on n-gram that allows word skipping. The calculation method of ROUGE-L is shown in Equation \ref{con:ROUGE-L},where $F_{LCS}$ is ROUGE-L and $\beta$ is a constant.
    
\begin{equation}
\begin{aligned}
Recall_{L C S} &=\frac{L C S(C, S)}{\operatorname{len}(S)} \\
Precision_{L C S} &=\frac{L C S(C, S)}{\operatorname{len}(C)} \\
F_{L C S} &=\frac{\left(1+\beta^{2}\right) Recall_{L C S} Precision_{L C S}}{Recall_{L C S}+\beta^{2} Precision_{L C S}}
\end{aligned}\label{con:ROUGE-L}
\end{equation}
    \item \textbf{METEOR} The Metric for Evaluation of Translation with Explicit ORdering (METEOR)\cite{SatanjeevBanerjee2005METEORAA} is a precision-based metric for the evaluation of machine-translation output. It overcomes some of the pitfalls of the BLEU score, such as exact word matching whilst calculating precision. The METEOR score allows synonyms and stemmed words to be matched with a reference word. The most important thing in meteor is to use WordNet thesaurus to align the generated sequence with the target sequence. 
    \item \textbf{Word Error Rate (WER)} It is important to substitute, delete, or insert some words into the generated sequence during the generation task in order to keep the generated sequence consistent with the target sequence. WER, which is defined as Equation \ref{con:WER}, can be used to assess the quality of the generated sequence.
\begin{equation}    \mathrm{WER}=100 \cdot \frac{Substitution+Deletion+Insertion}{N}\%\label{con:WER}
\end{equation}
    \item \textbf{CodeBLEU} CodeBLEU\cite{ShuoRen2020CodeBLEU} is a metric designed for code based on BLEU, which can pay attention to the keywords, leverage the tree structure and consider the semantic logic. It is defined as the weighted combination of $\mathrm{BLEU}$, $\mathrm{BLEU}_{\text {weight }}$,$\mathrm{Match}_{\text {ast}}$ and $\mathrm{Match}_{\text{df}}$, which is shown in Equation \ref{con:codebleu}. The first term refers to the standard $\mathrm{BLEU}$. $\mathrm{BLEU}_{\text {weight }}$ is the weighted n-gram match, obtained by comparing the hypothesis code and the reference code tokens with different weights. $\mathrm{Match}_{\text {ast}}$ is the syntactic AST match, exploring the syntactic information of code. $\mathrm{Match}_{\text{df}}$ is the semantic data-flow match, considering the semantic similarity between the prediction and the reference. The weighted n-gram match and the syntactic AST match are used to measure grammatical correctness, whereas the semantic data-flow match is used to calculate logic correctness. 
\begin{equation}
 \begin{aligned}
\text { CodeBLEU } &=\alpha \cdot \mathrm{BLEU}+\beta \cdot \mathrm{BLEU}_{\text {weightt }} \\
&+\gamma \cdot \mathrm{Match}_{\text {ast}}+\delta \cdot \mathrm{Match}_{\text {df}}
\end{aligned}
\label{con:codebleu}
\end{equation}
\end{itemize}

\subsubsection{Information Retrieval related Metrics}
Because some code-related tasks, such as Code Search, are similar to information retrieval, many code-related metrics are associated to information retrieval field.
\begin{itemize}
    \item \textbf{SuccessRate(SR)} If the information matching the input is in the top-k of the search information sorting list, the search is successful. SR@k Calculate the proportion of successful searches in all searches. The calculation method is shown in Equation \ref{con:SR}, where $\delta$ is a constant, $Frank_k$ is the rank of matched information in the search information sorting list and $Frank_q \leq K $ means successful retrieval.
    \begin{equation}
    \mathrm{SR@K}=\frac{1}{|Q|} \sum_{q=1}^{Q} \delta\left(\operatorname{FRank}_{q} \leq K\right) \label{con:SR}
    \end{equation}
    \item \textbf{Mean Reciprocal Rank (MRR)} In the retrieval information sorting list, MRR considers the ranking of the retrieved matching information. The score is $\frac{1}{N}$ if the nth information in the list fits the input, and $0$ if there is no matching sentence. The sum of all scores is the final score. Equation \ref{con:MRR} shows how to calculate MRR. 
    \begin{equation}
    \mathrm{MRR}=\frac{1}{|Q|} \sum_{q=1}^{|Q|} \frac{1}{\operatorname{FRank}_{q}}\label{con:MRR}
    \end{equation}
    \item \textbf{Best Hit Rank} Best Hit Rank is the highest rank of the hit snippets for the query. A highest best hit implies lower user effort to inspect the desired hit snippet.
\end{itemize}

\subsection{Datasets}

We summarize various datasets in the table \ref{table:datasets}. Some works collect and generate their own datasets for study, which may cause the difficulty on comparing different works on the same tasks. Table \ref{table:datasets} does not contain any datasets that are not open-source.

\begin{table}
\centering
\caption{Datasets}
\label{table:datasets}
\resizebox{\linewidth}{!}{%
\begin{tabular}{>{\centering\hspace{0pt}}m{0.298\linewidth}|>{\centering\hspace{0pt}}m{0.154\linewidth}|>{\hspace{0pt}}m{0.48\linewidth}} 
\hline\hline
Name              & Study    & Description                       \\ 
\hline
\href{https://github.com/xiaojunxu/dnn-binary-code-similarity}{Genius Dataset}           & \cite{QianFeng2016ScalableGB,Xu2017}       & A real-world dataset of 33,045 devices which was collected from public sources and our system                                  \\ 
\hline
\href{https://github.com/dakuo/haconvgnn}{notebookcdg} &\cite{Liu2021}  &Has 28,625 code–documentation pairs                                   \\ 
\hline
\href{http://www.srl.inf.ethz.ch/py150}{py150}             & \cite{Raychev2016,Kanade2019,SeohyunKim2020CodePB}   & Consists of parsed ASTs collected from GitHub python repositories by removing duplicate files \\ 
\hline
\href{http://www.srl.inf.ethz.ch/js150}{js150}            & \cite{Raychev2016,Wang2021}   & Consists of 150'000 JavaScript files and their corresponding parsed ASTs   \\ 
\hline
\href{https://github.com/shangqing-liu/CCSD-benchmark-for-code-summarization}{HGNN}            & \cite{Liu2020HybridGNN}   & Crawled from diversified large-scale open-source C projects (total 95k+ unique functions in the dataset)                                  \\ 
\hline
\href{https://github.com/github/CodeSearchNet}{CodeSearchNet}          & \cite{husain2019codesearchnet, Feng2020CodeBERTAP,Ugner}   & Consists of 2 million (comment, code) pairs from open source libraries                                    \\ 
\hline
\href{https://github.com/sriniiyer/concode}{CONCODE}           & \cite{SrinivasanIyer2018MappingLT,allamanis2019adverse}   & Consists over 100,000 examples consisting of Java classes from online code repositories                                   \\ 
\hline
\href{https://github.com/microsoft/CodeXGLUE}{CodeXCLUE}           & \cite{ShuaiLu2021CodeXGLUEAM,RuchirPuri2021CodeNetAL,YueWang2021CodeT5IU}   & Includes a collection of 10 tasks across 14 datasets                                   \\ 
\hline
\href{https://github.com/eth-sri/TFix}{TFix's Code Patches Data}            &\cite{BerkayBerabi2021TFixLT}   & Contains more than 100k code patch pairs extracted from open source projects on GitHub                                   \\ 
\hline
\href{https://github.com/csebuetnlp/CoDesc}{CoDesc}            & \cite{hasan2021codesc}  &  A large dataset of 4.2m Java source code and parallel data of their description from code search, and code summarization studies                                 \\ 
\hline
\href{https://bitbucket.org/iiscseal/deepfix}{DeepFix}           & \cite{RahulGupta2017DeepFixFC,Yasunaga2020,ZiminChen2021SEQUENCERSL}   &Consists of a program repair dataset (fix compiler errors in C programs)                                   \\ 
\hline
\href{https://github.com/michiyasunaga/DrRepair}{SPoC}         & \cite{Yasunaga2020}   &A program synthesis dataset, containing 18,356 programs with human-authored pseudocode and test cases                                   \\ 
\hline
\href{https://git.io/JJGwU}{Defects4J}         & \cite{Chakraborty2020}   & A large dataset of 32k real code change                                  \\ 
\hline
\href{http://leclair.tech/data/funcom/}{FunCom}            & \cite{AlexanderLeClair2019RecommendationsFD,PiyushShrivastava2021NeuralCS,BolinWei2019RetrieveAR,JunayedMahmud2021CodeTC}   &A collection of ~2.1 million Java methods and their associated Javadoc comments                                   \\ 
\hline
\href{https://github.com/Jun-jie-Huang/CoCLR}{CoSQA}            & \cite{JunjieHuang2021CoSQA2W,li2022coderetriever}   & Includes 20,604 labels for pairs of natural language queries and codes, each annotated by at least 3 human annotators                                   \\ 
\hline
\href{https://conala-corpus.github.io/}{CoNaLa}           & \cite{PengchengYin2018LearningTM,PengchengYin2018TRANXAT}   & Consists 2379 training and 500 test examples that were manually annotated                                  \\ 
\hline
\href{https://github.com/odashi/ase15-django-dataset}{Django}            & \cite{YusukeOda2015LearningTG,XiVictoriaLin2018NL2BashAC}   & Comprises of 16000 training, 1000 development and 1805 test annotations                                \\ 
\hline
\href{https://github.com/wala/blanca}{BLANCA}           & \cite{IbrahimAbdelaziz2021CanMR}   & A collection of benchmarks that assess code understanding based on tasks                                  \\ 
\hline
\href{https://github.com/indobenchmark/indonlg}{IndoNLG}           & \cite{SamuelCahyawijaya2021IndoNLGBA}   & A collection of Natural Language Generation (NLG) resources for Bahasa Indonesia with 6 kind of downstream tasks                                  \\ 
\hline
\href{https://github.com/facebookresearch/Neural-Code-Search-Evaluation-Dataset}{Neural-Code-Search-Evaluation-Dataset}           & \cite{HongyuLi2019NeuralCS}   & An evaluation dataset consisting of natural language query and code snippet pairs for code search                                 \\ 

\hline\hline
\end{tabular}
}
\end{table}

\section{Open Problems}
\label{open-problems}
It is difficult to fully capture the information of the structures and semantics of codes with the existing technology. Most deep-learning models are designed for specific tasks and a single language, which are lack flexibility. The following open problems can be considered as the future work direction.

\paragraph{\textbf{Information capturing}} 
Many approaches use structural information in code, however the majority of them just use structure information such as AST to capture syntax information. There are only a few approaches to learning the whole representation of code that combine structure and semantic information (such as DFG), and they are only applied to specific tasks, not all tasks. As a result, one of the goals of future study could be to improve the model's ability to use the structure and semantic information of codes in various tasks.
\paragraph{\textbf{Flexibility}} 
The methods we described above are suitable to a certain task or dataset and lack flexibility. The model's flexibility implies it may be utilized in a range of scenarios, including those using datasets from shorter programs (with small graph structures) or smaller sample sizes, as well as scenarios involving several downstream jobs or programming languages. As a result, future study could concentrate on how to train a model that can accommodate a variety of circumstances.

\paragraph{\textbf{Model Simplicity}} 
Recent models, particularly those based on Transformer and its derivatives, have improved performance, but they often need more effort and machine capability. Therefore, it is necessary to propose lightweight model under the premise of ensuring accuracy.

\paragraph{\textbf{Explainability}} 
Deep learning approaches have always had explainability issues, and the approaches for code are no exception. While the current usage of attentional mechanisms in code generation tasks can explain the origins of token generation, there are still no good explanations for other tasks like safety analysis. Simultaneously, the model's explainability is very useful in determining structural information and producing superior metrics in the code. As a result, additional research into the explainability of code models is still worthwhile.

\paragraph{\textbf{Robustness}}
The robustness of code-domain models has not yet been studied, but as model performance improves, the robustness of code-domain models is sure to become a hot study area in the future. In the code domain, both sequence-based and graph-based models rely on the representation of code tokens to some extent, which leads to model performance reduction when test code fragments are not represented in the same way as training code fragments (e.g., different representations of API calls in the python language). Furthermore, while code graph-based models can better capture the information of code fragments, graph structures are sensitive to attacks. There have been numerous techniques to explore the robustness of graph models, and how to convert them to work on code graphs is an important research field.

\paragraph{\textbf{Metrics}}
In the previous section, the metrics for evaluate the effectiveness of the code representation are based on the the Natural Language Processing area and Information Retrieval area. Although there are metrics designed specifically for code representation, there is a small amount of number of metrics proposed to suit the code data and downstream tasks. The following are the potential directions that can be studied:

\textbf{1) Measure of information} The requirement for appropriate measures for the information used in deep learning models is growing as the variety of structures employed in learning the representation for codes increases. Recent metrics that measure how models utilise this information in these structures are performed after the downstream tasks, however measures directly during the code representation stage have never been provided.

\textbf{2) Measure of Explainability} As previous mentioned, generating better metrics for measuring model efficacy on downstream tasks is vital for the explainability of models, which can  better qualify how the model works.

\textbf{3) Measure for Bias Problem} The code data will inevitably contains repeat and duplication, which might contribute to a bias problem. As far as we know, there have been few studies and discussions on the bias problem in code representation. Therefore, it is a new future direction to consider the bias in code data, while reasonable metrics for measuring  the bias of the models in code are required.

\section{Conclusion}
\label{conclusion}
It is critical to understand the structural and semantic meaning of codes when working on intelligent software. In this survey, We give a comprehensive overview of structure-based methods to code representation learning in recent years, which we divide into two groups: sequence-based models and graph-based models, then summarize and compare the methods in each group. The downstream tasks, as well as metrics and datasets, are also introduced here. It is shown that deep learning models are useful in code understanding, and further multiple downstream tasks. However, the field of code comprehension is still in its infancy, with numerous obstacles and unsolved issues. Finally, as future directions for code understanding, we offer four open questions.


\begin{acks}
\end{acks}



\appendix

\end{document}